\documentclass[letterpaper,11pt]{article}

\pdfoutput=1

\usepackage{newchet}

\hypersetup{pdftitle={The Gradient Structure of Perturbative RG Flows}}

\newcommand{\dd}{\mathrm{d}}
\newcommand{\ve}{\varepsilon}
\newcommand{\cN}{\mathcal N}

\newcommand{\cE}{\mathcal E}
\newcommand{\cJ}{\mathcal J}
\newcommand{\bh}{\hat\beta}
\newcommand{\Bh}{\hat B}
\NewDocumentCommand{\stL}{e{_}}{{\text{\fontsize{11pt}{0pt}\selectfont$\mathsterling$}}\IfValueT{#1}{_{\mkern-3mu#1}}}
\NewCommandCopy\nablasymbol\nabla
\RenewDocumentCommand{\nabla}{e{_}}{\nablasymbol\IfValueT{#1}{_{\mkern-5mu#1}}}

\renewcommand{\leq}{\leqslant}

\renewcommand{\geq}{\geqslant}
\newcommand{\lsp}{\hspace{0.5pt}}
\newcommand{\lnsp}{\hspace{-0.5pt}}
\newcommand{\anomfont}[1]{\mathsf{#1}}
\newcommand{\sA}{\anomfont{A}}
\newcommand{\sB}{\anomfont{B}}
\newcommand{\sC}{\anomfont{C}}
\newcommand{\sD}{\anomfont{D}}
\newcommand{\sE}{\anomfont{E}}
\newcommand{\sF}{\anomfont{F}}
\newcommand{\sG}{\anomfont{G}}
\newcommand{\sH}{\anomfont{H}}
\newcommand{\sI}{\anomfont{I}}
\newcommand{\sP}{\anomfont{P}}
\newcommand{\sS}{\anomfont{S}}
\newcommand{\sT}{\anomfont{T}}
\newcommand{\sU}{\anomfont{U}}
\newcommand{\sV}{\anomfont{V}}
\newcommand{\sw}{\anomfont{w}}
\newcommand{\sQ}{\anomfont{Q}}
\newcommand{\sJ}{\anomfont{J}}
\newcommand{\sK}{\anomfont{K}}

\renewcommand{\ss}{\anomfont{s}}
\renewcommand{\sf}{\anomfont{f}}
\usepackage[scr=boondoxo]{mathalpha}
\newcommand{\gs}{\mathscr g}
\usepackage[Euler]{upgreek}
\DeclareFontFamily{U}{eurs}{\skewchar\font'177}
\DeclareFontShape{U}{eurs}{m}{n}{<-6>s*[0.93]eurm5<6-8>s*[0.9]eurm7<8->s*[0.93]eurm10}{}
\DeclareFontShape{U}{eurs}{b}{n}{<-6>s*[0.93]eurb5<6-8>s*[0.9]eurb7<8->s*[0.93]eurb10}{}
\makeatletter
\SetSymbolFont{ugrf@m}{normal}{U}{eurs}{m}{n}
\SetSymbolFont{ugrf@m}{bold}{U}{eurs}{b}{n}
\makeatother
\renewcommand{\upsilon}{\upupsilon}

\DeclareMathOperator{\Rea}{\text{Re}}

\title{The Gradient Structure of Perturbative RG Flows}

\author{\texorpdfstring{Lorenzo Benfatto,\(^{\!a}\) William H.\ Pannell,\(^{\!b}\) and Andreas Stergiou\(^b\)}{Lorenzo Benfatto, William H. Pannell, and Andreas Stergiou}}

\affiliation{\(^a\)Universit\`a di Pisa and INFN - Sezione di Pisa, Largo Bruno Pontecorvo 3, 56127 Pisa, Italy\\[3pt]
\(^b\)Department of Mathematics, King's College London, Strand, London WC2R 2LS, United Kingdom}

\email{lorenzo.benfatto@phd.unipi.it, whp23@cantab.ac.uk, andreas.stergiou@kcl.ac.uk}

\abstract{Weyl consistency conditions show that perturbative renormalisation group flows in two, four and six dimensions are almost gradient flows. The obstruction is an antisymmetric contribution to the metric on the space of couplings, which has resisted removal since the work of Jack and Osborn more than 35 years ago. We show that a conformal Ward identity on the sphere allows it to be removed by a choice of scheme. The beta function is then the gradient of a potential $A$ with respect to a symmetric metric, to all orders in perturbation theory. The Ward identity gives this result in any dimension $d$, in theories with and without a Weyl anomaly. At leading order the metric is fixed by the two-point functions of the perturbing operators, so unitarity makes it positive definite at weak coupling. At fixed points $A$ equals $\widetilde{F}=\sin\tfrac{\pi d}{2}\,\ln Z_{S^{d}}$, so the $\widetilde{F}$-theorem follows for perturbative flows to all loop orders. We show that sources of relevant operators, which on the sphere mix with the curvature, do not affect these results. In unitary theories the only exception is the source of a free scalar field near $d=6,10,\ldots\,$, so that for $\phi^3$ theory near six dimensions the gradient property remains open.}

\date{November 2026}

\begin{document}

\maketitle

\tableofcontents

\section{Introduction}\label{sec:intro}
The renormalisation group (RG) has proven to be one of the most profound developments in the study of quantum field theory (QFT). At its most basic level, the RG appears as a dynamical system governing how the coupling constants $g^I$ of a theory change as the result of a change in energy scale $\mu$,
\begin{equation}
    \mu\frac{\dd g^I}{\dd\mu}=\beta^I(g)\,.
\end{equation}
Wilson recognised early on that equations of this form can in principle show complicated and exotic behaviour \cite{Wilson:1971bg}. In unitary QFTs, however, such behaviour does not appear to arise. By the maxim that whatever is not forbidden is realised in some example, this suggests a deeper organising principle. In 1974, Wallace and Zia proposed such a mechanism for perturbative RG \cite{Wallace:1974dx,Wallace:1974dy}, noting that through three loops the beta function of multiscalar models with a quartic potential in four dimensions can be written as the gradient of a scalar function $A$,
\begin{equation}\label{eq:gradient flow}
    \beta^I(g)=G^{IJ}(g)\lsp\partial_J A(g)\,,
\end{equation}
where $G_{IJ}$ acts as a Riemannian metric on the space of couplings. Dynamical systems governed by a gradient vector field are exceptionally simple. Their solutions either move from one fixed point to another or run off to infinity \cite{lojasiewicz1982trajectoires}. In particular, $A$ decreases monotonically towards the IR whenever $G_{IJ}$ is positive definite,
\begin{equation}\label{eq:strong monotonicity}
    \mu\frac{\dd A}{\dd\mu}=\beta^IG_{IJ}\beta^J\geq0\,.
\end{equation}

Three strengths of monotonicity are usually distinguished. Weak monotonicity demands a quantity $A$, defined for conformal field theories (CFTs), such that $A_{\text{UV}}>A_{\text{IR}}$ whenever an RG flow connects two CFTs. Strong monotonicity demands that $A$ be defined along the entire flow and decrease towards low energies, $\mu\,\dd A/\dd\mu\geq0$. Full gradient flow demands a solution of (\ref{eq:gradient flow}) with a symmetric and positive-definite metric $G_{IJ}$. It implies that $A$ is a strong monotone, and also that the eigenvalues of the stability matrix of the beta function are real at fixed points. These eigenvalues are $\Delta-d$, where $\Delta$ are the scaling dimensions of the nearly marginal operators at the fixed point. Full gradient flow has not yet been established in general.

In perturbation theory, the local renormalisation group has partly extended the conclusions of Wallace and Zia to all loop orders. In four dimensions, Jack and Osborn \cite{Jack:1990eb} and Osborn \cite{Osborn:1991gm} showed that conditions on finiteness of correlation functions and Weyl consistency conditions independently imply
\begin{equation}\label{eq:almost gradient}
    \partial_I A=(G_{IJ}+\partial_I\sw_J-\partial_J\sw_I)\beta^J\,,
\end{equation}
where $\sw_I$ is a one-form on the space of couplings. This result was extended to two dimensions, in connection with the $c$-theorem \cite{Osborn:1991gm}, and to six dimensions \cite{Grinstein:2013cka,Stergiou:2016uqq}. Contracting with $\beta^I$ removes the antisymmetric part $\partial_I\sw_J-\partial_J\sw_I$ and gives the strong monotonicity condition (\ref{eq:strong monotonicity}). The antisymmetric part $\partial_I\sw_J-\partial_J\sw_I$, however, spoils a gradient interpretation of (\ref{eq:almost gradient}). Nevertheless, considerable evidence has emerged from finite-loop order studies of the gradient condition in a number of dimensions that $\sw_I$ does not obstruct gradient flow \cite{Gracey:2015fia,Jack:2018oec,Pannell:2024sia,Pannell:2025curvature,Pannell:2025ajf,Benfatto:2025yci,Benfatto:2026esr,Jack:2026npe}, and the gradient property seems to hold in more exotic perturbative settings as well~\cite{Jack:2026toa}. For conformal manifolds in two dimensions, an analysis of the three-point function $\langle T_{\mu\nu}\mathcal{O}_I\mathcal{O}_J\rangle$ of the energy-momentum tensor $T_{\mu\nu}$ and the marginal operators $\mathcal{O}_I$ shows that $\partial_I\sw_J-\partial_J\sw_I=0$ along the marginal directions \cite{Gomis:2015yaa}.

Our main result is that a conformal Ward identity on the sphere enforces a gradient condition for perturbative RG flows,
\begin{equation}
    \partial_IA=G_{IJ}\Bh^J\,,
\end{equation}
with a manifestly symmetric metric $G_{IJ}$ that is positive definite at weak coupling. Here $\Bh^I=\bh^I-(\upsilon g)^I$ is the beta function $\bh^I$ in $d$ dimensions, shifted to account for broken flavour currents \cite{Jack:1990eb,Osborn:1991gm,Jack:2013sha,Fortin:2012hn,Baume:2014rla,Herren:2021ambiguities}. It takes the place of $\beta^I$ in (\ref{eq:almost gradient}). The identity exists because the sphere has conformal Killing vectors that are not isometries. A Weyl rescaling by an $\ell=1$ spherical harmonic is undone by a diffeomorphism, and comparing the two transformations gives a condition on $\ln Z$, the logarithm of the sphere partition function, that goes beyond the Weyl consistency conditions. Osborn and Shore studied correlation functions of the energy-momentum tensor on spaces of constant curvature, with a view to a four-dimensional analogue of the $c$-theorem \cite{Osborn:1999az}. They found that the two-point function at separated points is not directly related to the conformal anomaly $a$. Our identity involves instead the correlation functions of the operators $\mathcal{O}_I$.

Away from even dimensions there is no Weyl anomaly, and the Ward identity gives the gradient condition directly, with
\begin{equation}
    A=\sin\frac{\pi d}{2}\left(1+\frac{1}{d}\Bh^I\partial_I\right)\ln Z\,,\qquad G_{IJ}=\frac{V_d}{d}\sin\frac{\pi d}{2}\,(H^{(0)}-H^{(1)})_{IJ}\,,
\end{equation}
where $V_d$ is the volume of the unit sphere $S^d$ and $H^{(\ell)}_{IJ}$ are the angular moments of the Hessian $\delta^2\ln Z/\delta g^I\delta g^J$. Near an even dimension $D$ we work in $d=D-\varepsilon$ and require all quantities to be regular as $\varepsilon\to0$. The Weyl anomaly then adds a one-form to the Ward identity. The response of the identity to a change of the radius $r$ of the sphere removes this one-form, and we find
\begin{equation}
    A=\frac{\sin\frac{\pi d}{2}}{\varepsilon}\left(1+\frac{1}{d}\Bh^I\partial_I\right)(r\partial_r+\varepsilon)\ln Z\,,\qquad G_{IJ}=\frac{\sin\frac{\pi d}{2}}{d\lsp\varepsilon}(r\partial_r+\varepsilon)[V_d(H^{(0)}-H^{(1)})]_{IJ}\,,
\end{equation}
with $V_d$ now the volume of the sphere of radius $r$. These expressions are regular in $\varepsilon$, so the limit $\varepsilon\to0$ gives a gradient condition in $D$ dimensions. In the language of the local renormalisation group, the sphere Ward identity guarantees a scheme in which $\sw_I=0$ in (\ref{eq:almost gradient}), as we show in sections \ref{sec:twod} and \ref{sec:fourd}.

At fixed points the potential equals the rescaled sphere free energy of Giombi and Klebanov \cite{Giombi:2014xxa}, which was used in the $\varepsilon$ expansion in \cite{Fei:2015oha},
\begin{equation}
    A_*=\widetilde{F}_*\,,\qquad \widetilde{F}_*=\sin\frac{\pi d}{2}\,\ln Z_{\text{CFT}}\,.
\end{equation}
Here $Z_{\text{CFT}}$ is the sphere partition function of the fixed-point CFT in the convention of these references, in which it does not depend on the radius. Away from even dimensions it is the partition function $Z$ at the fixed point. Near an even dimension $D$, $Z_{\text{CFT}}$ and $Z$ differ. The renormalised $\ln Z$ is regular as $\varepsilon\to0$ and depends on the radius through the Weyl anomaly, while $\ln Z_{\text{CFT}}$ has a pole in $\varepsilon$, which the sine cancels. For even $D$ the potential at fixed points becomes the integrated Weyl anomaly $\mathcal{A}$ on the sphere,
\begin{equation}
    A_*=\frac{(-1)^{\frac{D}{2}+1}\pi}{2}\mathcal{A}_*\,,
\end{equation}
which is the quantity that Cardy conjectured to be monotonic \cite{Cardy:1988cwa}. At leading order in the couplings, $G_{IJ}$ is a positive multiple of the Zamolodchikov metric, and is therefore positive definite in a perturbative region around each unitary CFT. For flows that stay within such regions, we thus establish a weak version of the $\widetilde{F}$-theorem to all loop orders,
\begin{equation}
    \widetilde{F}_{\text{UV}}-\widetilde{F}_{\text{IR}}=\int_0^\infty\dd\mu\lsp\frac{\dd A}{\dd \mu}=\int_0^\infty\frac{\dd\mu}{\mu}\,\Bh^IG_{IJ}\Bh^J\geq0\,.
\end{equation}

The results stated so far come with a caveat. On the sphere, the sources of relevant operators cannot simply be set to zero, because they mix with the curvature and with derivatives of the couplings.\footnote{This concerns sources whose scaling dimension is an even integer. Factors of the curvature and pairs of derivatives add an even number of dimensions. The Weyl variation of a relevant source whose scaling dimension is not an even integer therefore vanishes when all such sources vanish, so these sources can consistently be set to zero.} They add a term to the Ward identity, which could spoil the gradient condition. We show that this term disappears when the sources are fixed by a stationarity condition. Without a Weyl anomaly, the condition states that the one-point functions of the relevant operators vanish on the sphere. The gradient condition then holds with a potential and a metric from which the sources have been eliminated. The elimination changes neither the leading term in the metric nor the value of $A$ at fixed points, but it requires the second derivatives of the sphere free energy with respect to the sources to form an invertible matrix. This is the case near odd dimensions, and near an even dimension $D$ for sources of scaling dimension at most $D/2$, such as the mass sources near four dimensions. In unitary theories the only exception is the source of a free scalar field near $D=6,10,\ldots\,$. In $\phi^3$ theory near six dimensions the source of the elementary scalar is of this type, and whether the gradient condition holds there remains an open question, although perturbative evidence \cite{Gracey:2015fia,Benfatto:2025yci} suggests a positive answer in that case too.

This paper is organised as follows. Section \ref{sec:localRG} reviews the local renormalisation group. Sections \ref{sec:free} and \ref{sec:anom} use the sphere Ward identity to prove the gradient condition in theories without and with a Weyl anomaly. Section \ref{sec:relsource} restores the sources of relevant operators and shows how they are eliminated. Section \ref{sec:specificd} connects our results to earlier work on the local renormalisation group in $d=2$, $4$ and $6$, and section \ref{sec:conclusion} contains our conclusions. Appendix \ref{app:pertchecks} checks the gradient property against explicit beta functions of $\phi^4$ theories, and appendix \ref{app:consistency} derives the Weyl consistency conditions in $d=4-\varepsilon$.

Throughout, we work in perturbation theory. All quantities are formal perturbative series, which we assume to exist and to be regular at each order. By regular we mean that, after renormalisation, no poles appear either in $\varepsilon$ or on approaching a fixed point. We make no claim about the convergence of these series. We also assume that all counterterms, anomalies and source transformations are local. Gauge couplings require two qualifications, which we discuss in section \ref{sec:fourd}.

\paragraph{Note added.} While this paper was in the final stages of preparation, we learned about work by Marten Reehorst \cite{Reehorst:2026toa1,Reehorst:2026toa2} that may have some overlap with our results.

\section{The local renormalisation group}\label{sec:localRG}
We consider a renormalisable theory of fields $\phi$ (not necessarily scalars) with a classically conformally invariant Euclidean action $S[\phi]$. The action contains classically dimensionless couplings $g^I$. We assume that it is also defined for position-dependent couplings $g^I(x)$ and for a general background metric $\gamma_{\mu\nu}(x)$. The partition function is
\begin{equation}\label{eq:partitionFunction}
    Z(\gamma,g)=\int \mathscr{D}\phi\, e^{-S}\,.
\end{equation}
The inverse metric $\gamma^{\mu\nu}$ sources the energy-momentum tensor $T_{\mu\nu}$ and the couplings $g^I$ source corresponding classically marginal operators $\mathcal{O}_I$. These operators are defined by
\begin{equation}\label{eq:TmnOidefs}
    T_{\mu\nu}(x)=\frac{2}{\sqrt{\gamma(x)}} \frac{\delta S}{\delta \gamma^{\mu\nu}(x)}\,, \qquad \mathcal{O}_I(x)=\frac{1}{\sqrt{\gamma(x)}}\frac{\delta S}{\delta g^I(x)}\,,
\end{equation}
where $\gamma(x)=\det(\gamma_{\mu\nu}(x))$. As a consequence of \eqref{eq:partitionFunction} and \eqref{eq:TmnOidefs} we have
\begin{equation}\label{eq:1ptTO}
    \begin{aligned}
        \langle T_{\mu\nu}(x)\rangle &=\frac{1}{Z(\gamma,g)}\int \mathscr{D}\phi\, T_{\mu\nu}(x)\lsp e^{-S}=-\frac{2}{\sqrt{\gamma(x)}} \frac{\delta \ln Z}{\delta \gamma^{\mu\nu}(x)}\,,\\
        \langle\mathcal{O}_I(x)\rangle &=\frac{1}{Z(\gamma,g)}\int \mathscr{D}\phi\, \mathcal{O}_I(x)\lsp e^{-S}=-\frac{1}{\sqrt{\gamma(x)}}\frac{\delta \ln Z}{\delta g^I(x)}\,.
    \end{aligned}
\end{equation}

With a general metric and position-dependent couplings, the counterterms in $S$ include derivatives of the metric and of the couplings. They make insertions of $T_{\mu\nu}$ and $\mathcal{O}_I$ finite in correlation functions such as \eqref{eq:1ptTO}, and also in higher-point functions~\cite{Jack:1990eb}. We define connected $n$-point functions, including contact terms, as
\begin{equation}\label{eq:Tnpt}
    \langle T_{\mu_1\nu_1}(x_1)\ldots T_{\mu_n\nu_n}(x_n)\rangle=\frac{(-2)^n}{\sqrt{\gamma(x_1)}\cdots \sqrt{\gamma(x_n)}}\frac{\delta}{\delta \gamma^{\mu_1\nu_1}(x_1)}\cdots \frac{\delta}{\delta \gamma^{\mu_n\nu_n}(x_n)}\ln Z\,,
\end{equation}
and
\begin{equation}
    \langle \mathcal{O}_{I_1}(x_1)\ldots \mathcal{O}_{I_n}(x_n)\rangle=\frac{(-1)^n}{\sqrt{\gamma(x_1)}\cdots \sqrt{\gamma(x_n)}}\frac{\delta}{\delta g^{I_1}(x_1)}\cdots \frac{\delta}{\delta g^{I_n}(x_n)}\ln Z\,.
\end{equation}
These definitions are symmetric under reordering of the operators. The renormalised flat-space correlators follow by setting, after all variations, the couplings to constants and the metric to $\delta_{\mu\nu}$. For functions of the couplings, $\partial_I$ denotes the derivative with respect to $g^I$. On functionals such as $\ln Z$ we define $\partial_I=\int\dd^d x\,\delta/\delta g^I(x)$, which reduces to the ordinary derivative at constant couplings.

The theory at $g^I=0$ may have a flavour symmetry group $G$. The couplings then transform in a representation of $G$, and theories whose couplings are related by such a transformation are equivalent. We write an infinitesimal flavour transformation as $\delta g^I=-(\omega g)^I$, with $\omega$ in the Lie algebra of $G$. The associated currents $J^\mu$ are conserved at $g^I=0$, where they have dimension exactly $d-1$. We couple them to a background gauge field $a_\mu$, valued in the same Lie algebra, so that the partition function becomes $Z(\gamma,g,a)$. In analogy with \eqref{eq:TmnOidefs} and \eqref{eq:1ptTO},
\begin{equation}\label{eq:Jdefn}
    J^\mu(x)=\frac{1}{\sqrt{\gamma(x)}}\frac{\delta S}{\delta a_\mu(x)}\,,\qquad \langle J^\mu(x)\rangle=-\frac{1}{\sqrt{\gamma(x)}}\frac{\delta\ln Z}{\delta a_\mu(x)}\,.
\end{equation}
Flavour transformations can now be made local. With $\omega=\omega(x)$, the sources transform as
\begin{equation}
    \delta g^I=-(\omega g)^I\,,\qquad \delta a_\mu=D_\mu\omega=\partial_\mu\omega+[a_\mu,\omega]\,,
\end{equation}
so that $D_\mu g^I=\partial_\mu g^I+(a_\mu g)^I$ transforms covariantly. We assume that the regularisation and the counterterms respect this symmetry. Then $\ln Z$ is invariant, and
\begin{equation}\label{eq:flavourWard}
    \int\dd^d x\left[-(\omega g)^I\frac{\delta}{\delta g^I}+D_\mu\omega\cdot\frac{\delta}{\delta a_\mu}\right]\ln Z(\gamma,g,a)=0\,,
\end{equation}
where the dot denotes contraction of Lie algebra indices. At $a_\mu=0$ this flavour Ward identity states that $\omega\cdot\nabla_\mu\langle J^\mu\rangle=-(\omega g)^I\langle\mathcal{O}_I\rangle$. The divergence of the currents is thus fixed by the flavour variation of the couplings. Here we have assumed that the flavour symmetry is not anomalous. In even $d$ it can be, and the right-hand side of \eqref{eq:flavourWard} is then a local functional of the sources that is linear in $\omega$.

The local renormalisation group describes the response of $\ln Z$ to a Weyl rescaling $\gamma_{\mu\nu}\to e^{2\sigma}\gamma_{\mu\nu}$ with arbitrary $\sigma(x)$~\cite{Jack:1990eb,Osborn:1991gm}. For constant $\sigma$ a Weyl rescaling is equivalent to a change of the renormalisation scale, which is compensated by the flow of the couplings. For local $\sigma$ the compensating changes of the sources are also local. To first order in $\sigma$,
\begin{equation}\label{eq:localRGbeta}
    \int\dd^d x\left[-2\sigma\gamma^{\mu\nu}\frac{\delta}{\delta\gamma^{\mu\nu}}+\sigma\bh^I\frac{\delta}{\delta g^I}+\left(\sigma\rho_I D_\mu g^I-\partial_\mu\sigma\,\upsilon\right)\cdot\frac{\delta}{\delta a_\mu}\right]\ln Z(\gamma,g,a)=\mathcal{A}_\sigma\,,
\end{equation}
where $\bh^I=\mu\,\dd g^I/\dd\mu$ is the beta function in $d$ dimensions, including its classical part. The one-form $\rho_I$ and the matrix $\upsilon$, which we call the beta shift, are functions of the couplings with values in the Lie algebra of $G$. The Weyl anomaly $\mathcal{A}_\sigma$ is a local functional of the sources, linear in $\sigma$. Here we have omitted sources for relevant operators other than $a_\mu$, which we discuss in section~\ref{sec:relsource}.

The flavour Ward identity removes the term with $\partial_\mu\sigma$. The matrix $\upsilon$ is built covariantly from the couplings, so $D_\mu\upsilon=\partial_I\upsilon\,D_\mu g^I$. Choosing $\omega=\sigma\upsilon$ in \eqref{eq:flavourWard} then gives
\begin{equation}
    \int\dd^d x\,\partial_\mu\sigma\,\upsilon\cdot\frac{\delta\ln Z}{\delta a_\mu}=\int\dd^d x\,\sigma\left[(\upsilon g)^I\frac{\delta}{\delta g^I}-\partial_I\upsilon\,D_\mu g^I\cdot\frac{\delta}{\delta a_\mu}\right]\ln Z\,.
\end{equation}
Substituting this into \eqref{eq:localRGbeta}, we obtain
\begin{equation}\label{eq:localRG}
    \Delta_\sigma\ln Z(\gamma,g,a)=\int\dd^d x\left[-2\sigma\gamma^{\mu\nu}\frac{\delta}{\delta\gamma^{\mu\nu}}+\sigma\Bh^I\frac{\delta}{\delta g^I}+\sigma\tilde{\rho}_I D_\mu g^I\cdot\frac{\delta}{\delta a_\mu}\right]\ln Z(\gamma,g,a)=\mathcal{A}_\sigma\,,
\end{equation}
with
\begin{equation}\label{eq:Bdefn}
    \Bh^I=\bh^I-(\upsilon g)^I\,,\qquad \tilde{\rho}_I=\rho_I+\partial_I\upsilon\,.
\end{equation}
This is the form of the local renormalisation group equation that we use in the rest of the paper~\cite{Fortin:2012hn,Jack:2013sha}. The functions $\Bh^I$ and $\bh^I$ differ by an infinitesimal flavour rotation of the couplings. Flavour-invariant functions of the couplings therefore change in the same way under both. The conformal fixed points are the zeros of $\Bh^I$~\cite{Fortin:2012hn,Luty:2012ww}. If the flavour symmetry is anomalous, the right-hand side of \eqref{eq:flavourWard} with $\omega=\sigma\upsilon$ is a local functional linear in $\sigma$, and we absorb it into the Weyl anomaly $\mathcal{A}_\sigma$.

In the rest of the paper we evaluate \eqref{eq:localRG} on the round sphere $S^d$. The sphere has non-constant conformal Killing vectors, so a Weyl rescaling by a suitable $\sigma$ changes the metric in the same way as a diffeomorphism. Subtracting the two identities removes the variation of the metric, leaving behind a Ward identity that involves only the other sources. This sphere Ward identity is the main tool of this paper.

\section{Theories without a Weyl anomaly}\label{sec:free}
We begin with theories without a Weyl anomaly, where the sphere Ward identity gives the gradient condition directly. Weyl anomalies arise in even dimensions, so in this section we take $d\notin2\mathbb{N}$. We consider a QFT on the unit sphere $S^d$, with round metric $\gamma_{\mu\nu}$ and nearly marginal couplings $g^I$. We include the background gauge field $a_\mu$ for the flavour currents $J^\mu$, which have dimension exactly $d-1$ at $g^I=0$. The sources of all other relevant operators, e.g.\ scalars of dimension close to $d-2$, are set to zero. Such relevant operators matter on the sphere, because the non-zero Ricci scalar mixes them with operators of dimension close to $d$. We examine their effects in section \ref{sec:relsource}.

The derivation of the sphere Ward identity uses spherical harmonics, the eigenfunctions of the Laplacian,
\begin{equation}
    \nabla^2 Y_\ell(x)=-\ell(\ell+d-1)Y_\ell(x)\,.
\end{equation}
In particular, a real $\ell=1$ harmonic $Y_1(x)=Y(x)$ satisfies
\begin{equation}\label{eq:Yprops}
\begin{gathered}
\nabla_\mu\partial_\nu Y=-Y \gamma_{\mu\nu}\,,\qquad \nabla^2 Y=-d Y\,,\\
    \int_{S^d}\dd v\, Y=0\,,\qquad \int_{S^d}\dd v\, Y^2= N_Y\,,\qquad \int_{S^d}\dd v\, \partial^\mu Y\partial_\mu Y=d N_Y\,,
\end{gathered}
\end{equation}
where $\dd v=\dd^d x\sqrt{\gamma}$ is the volume element and the last integral follows by integration by parts. We define
\begin{equation}\label{eq:HIJdefn}
    H_{IJ}(x,y)=\frac{1}{\sqrt{\gamma(x)} \sqrt{\gamma(y)}}\frac{\delta^2 \ln Z}{\delta g^I(x)\,\delta g^J(y)}\,,
\end{equation}
which by the definitions of section \ref{sec:localRG} is the connected correlation function $\langle\mathcal{O}_I(x)\mathcal{O}_J(y)\rangle$ of operators sourced by the couplings, with contact terms included. The commutativity of functional derivatives guarantees the symmetry of this tensor, while at constant couplings the rotational invariance of the system\footnote{For $d=1$, here and below, rotational invariance must be supplemented by reflection symmetry.} allows us to define its angular moments $H^{(\ell)}_{IJ}$ via
\begin{equation}\label{eq:angmoms}
   \int_{S^d}\dd v(y)\,H_{IJ}(x,y) Y_\ell(y)= H^{(\ell)}_{IJ}Y_\ell(x)\,.
\end{equation}
The zeroth harmonic $Y_0(x)=1$ relates the zeroth moment to the second derivative of $\ln Z$ at constant couplings,
\begin{equation}
    V_d H^{(0)}_{IJ}= \int_{S^d} \dd v(x)\, \dd v(y)\, H_{IJ}(x,y)=\partial_I\partial_J \ln Z\,,
\end{equation}
where $V_d=2\pi^{\frac{d+1}{2}}/\Gamma(\frac{d+1}{2})$ is the volume of the unit sphere $S^d$.

By the first relation in (\ref{eq:Yprops}), the Lie derivative of the metric along the vector $\xi_\mu=-\partial_\mu Y$ is
\begin{equation}
\stL_\xi\lsp\gamma_{\mu\nu}=\nabla_\mu\lsp\xi_\nu+\nabla_\nu\lsp\xi_\mu=2Y\gamma_{\mu\nu}\,,
\end{equation}
so that $\xi^\mu$ is a conformal Killing vector of $\gamma_{\mu\nu}$ with the non-constant conformal factor $\sigma=Y$. Invariance under the transformation generated by $\xi^\mu$ can then be expressed in two ways. First, as a diffeomorphism of the sphere, it gives
\begin{equation}\label{eq:diffidentity}
    \int_{S^d} \dd^d x\left[-2 Y\gamma^{\mu\nu}\frac{\delta}{\delta \gamma^{\mu\nu}}+\stL_\xi\lsp g^I\frac{\delta}{\delta g^I}+\stL_\xi\lsp a_\mu\cdot\frac{\delta}{\delta a_\mu}\right]\ln Z(\gamma,g,a)=0\,.
\end{equation}
Second, as a Weyl rescaling by $\sigma=Y$, it gives
\begin{equation}
    \Delta_Y \ln Z(\gamma,g,a)=\int_{S^d} \dd^d x\left[-2Y\lsp \gamma^{\mu\nu}\frac{\delta}{\delta \gamma^{\mu\nu}}+Y\Bh^I\frac{\delta}{\delta g^I}+Y\tilde{\rho}_I D_\mu g^I\cdot\frac{\delta}{\delta a_\mu}\right]\ln Z(\gamma,g,a)=0\,,
\end{equation}
where the right-hand side vanishes because the theory has no anomaly. Subtracting the two equations gives the sphere Ward identity,
\begin{equation}\label{eq:sphereWard}
    \int_{S^d} \dd^d x\left[(Y \Bh^I+\partial^\mu Y\lsp\partial_\mu g^I)\frac{\delta \ln Z(\gamma,g,a)}{\delta g^I}+(Y\tilde{\rho}_I D_\mu g^I-\stL_\xi\lsp a_\mu)\cdot\frac{\delta \ln Z(\gamma,g,a)}{\delta a_\mu}\right]=0\,,
\end{equation}
which must hold for any choice of the couplings $g^I(x)$ and of $a_\mu$. We will only use it at $a_\mu=0$, where the term with $\stL_\xi\lsp a_\mu$ vanishes.

The identity has useful consequences only for position-dependent couplings. Its utility for our purposes is revealed when we perturb constant couplings by the harmonic $Y$ itself,
\begin{equation}
    g^I(x)=g^I+z^I \lsp Y(x)\,,
\end{equation}
where $z^I$ is a small constant vector. Linearising the Ward identity in $z$, using (\ref{eq:Yprops}) and the vanishing of $\delta\ln Z/\delta a_\mu\propto\langle J^\mu\rangle$ at constant couplings, which follows from rotational symmetry, gives
\begin{equation}\label{eq:linWard}
    \frac{N_Y}{V_d}\lsp z^I\left[(d\lsp\delta_I{}^J+\partial_I\Bh^J)\,\partial_J\ln Z+V_d H^{(1)}_{IJ}\Bh^J\right]+\text{O}(z^2)=0\,.
\end{equation}
Besides the conformal Killing vector, we have used the compactness of the sphere, which removes boundary terms. We have also used the form (\ref{eq:localRG}) of the local renormalisation group equation. Its $a_\mu$ term has coefficient $Y\tilde\rho_I D_\mu g^I$, so at first order in $z$ it multiplies $\langle J^\mu\rangle$ at constant couplings, which vanishes. In the form (\ref{eq:localRGbeta}) there would be an extra term $-\partial^\mu Y\,\upsilon\cdot\delta\ln Z/\delta a_\mu$. This term survives at first order in $z$ through $\langle J^\mu\mathcal{O}_J\rangle$ and spoils the gradient form. The gradient property therefore holds for $\Bh^I$ and in general fails for $\bh^I$, in agreement with perturbative checks (see appendix \ref{app:pertchecks}). To put (\ref{eq:linWard}) in a gradient form, we rewrite
\begin{equation}
    \partial_I \Bh^J\partial_J \ln Z=\partial_I(\Bh^J\partial_J \ln Z)-V_d H_{IJ}^{(0)}\Bh^J\,,
\end{equation}
to obtain
\begin{equation}\label{eq:NAgrad}
    \partial_IA=G_{IJ}\Bh^J\,,\qquad A=\left(1+\frac{1}{d}\Bh^I\partial_I\right)\widetilde{F}\quad\text{and}\quad G_{IJ}=\frac{V_d}{d}\sin\frac{\pi d}{2}\,(H^{(0)}-H^{(1)})_{IJ}\,.
\end{equation}
The metric $G_{IJ}$ is manifestly symmetric, since $H_{IJ}$ is. We have multiplied by $\frac{1}{d}\sin\frac{\pi d}{2}$, which makes the metric positive definite at weak coupling, as we show below.\footnote{This factor vanishes for $d\in2\mathbb{N}$, where the theory has a Weyl anomaly. These dimensions are treated in section \ref{sec:anom}.} Following \cite{Giombi:2014xxa,Fei:2015oha}, we have also defined
\begin{equation}
    \widetilde{F}=\sin\frac{\pi d}{2}\,\ln Z\,,
\end{equation}
where $Z$ is the partition function on $S^d$.

The potential $A$ decreases towards the IR if $G_{IJ}$ is positive definite. To check this at weak coupling, let us fix a point $x$ and let $\theta$ be the angle between $x$ and $y$. The function $Y(y)=\cos\theta$ is an $\ell=1$ harmonic with $Y(x)=1$. Then,
\begin{equation}\label{eq:Gchordal}
\begin{split}
    G_{IJ}&=\frac{V_d}{d}\sin\frac{\pi d}{2}\int_{S^d}\dd v(y)\,(1-\cos\theta)H_{IJ}(x,y)\\
    &=\frac{1}{2d}\sin\frac{\pi d}{2}\int_{S^d}\dd v(x)\,\dd v(y)\,s_{xy}^2H_{IJ}(x,y)\,,
\end{split}
\end{equation}
where we have defined the chordal distance
\begin{equation}
    s_{xy}=2\sin\frac{\theta}{2}\,.
\end{equation}
We only demand positivity at weak coupling. By assumption, $G_{IJ}$ is a regular perturbative series in $g^I$, so that $G_{IJ}=G_{IJ}|_{g=0}+\text{O}(g)$. If $G_{IJ}|_{g=0}$ is positive definite, then $G_{IJ}$ computed to any finite order in the couplings is positive definite in a sufficiently small ball around the CFT at $g^I=0$, which we call the reference CFT. When the perturbing operators are close enough to marginal, this ball contains the weakly coupled fixed points and the flows between them, so that the gradient condition applies to these flows.

At $g^I=0$ the theory is conformal, and at non-coincident points $H_{IJ}(x,y)$ is the conformal two-point function of the perturbing operators $\mathcal{O}_I$,
\begin{equation}\label{eq:Hg0}
    H_{IJ}(x,y)|_{g=0}=\frac{C_{IJ}}{(s_{xy}^2)^{\Delta_I}}\,,
\end{equation}
where the two-point function coefficients $C_{IJ}$ vanish between operators of different dimensions, and near marginality means that $\Delta_I$ is close to $d$. Contact terms may still occur. The integral (\ref{eq:Gchordal}) diverges at $x=y$ for $d>2$, so it is defined only once the contact terms in $H_{IJ}(x,y)$ are specified. Contact terms proportional to $\delta(x,y)$ do not contribute, since $s_{xy}^2$ vanishes at $x=y$, but contact terms with derivatives of $\delta(x,y)$ must be accounted for.

The Ward identity (\ref{eq:sphereWard}) fixes these terms. We now show that it selects one extension of (\ref{eq:Hg0}) to coincident points, namely the one defined by analytic continuation in $\Delta_I$. Near $g^I=0$, $\Bh^I=-(d-\Delta_I)\lsp g^I+\text{O}(g^2)$, while $\ln Z$ has the expansion
\begin{equation}
    \ln Z(\gamma,g)=\ln Z(\gamma,0)+\frac{1}{2}\int_{S^d} \dd v(x)\,\dd v(y)\,g^{I}(x)\lsp\big(H_{IJ}(x,y)|_{g=0}\big)\lsp g^J(y)+\text{O}(g^3)\,,
\end{equation}
where we have set $a_\mu=0$ and used that the Ward identity forbids a linear term. At second order in the couplings, (\ref{eq:sphereWard}) states that the quadratic term
\begin{equation}\label{eq:Hinvint}
    \frac{1}{2}\int_{S^d} \dd v(x)\,\dd v(y)\,g^{I}(x)\lsp \big(H_{IJ}(x,y)|_{g=0}\big)\lsp g^J(y)
\end{equation}
is invariant under the transformation\footnote{There is also a term from $\delta a_\mu=Y\tilde\rho_I\partial_\mu g^I$ at $\text{O}(g^2)$, proportional to the $\text{O}(g)$ term of $\langle J^\mu\rangle$. It vanishes by the flavour Ward identity.}
\begin{equation}\label{eq:gtransform}
    \delta g^I=-(d-\Delta_I)\lsp Y g^I+\partial^\mu Y\lsp\partial_\mu g^I\,,
\end{equation}
for any $g^I(x)$. To pick out specific moments of $H_{IJ}$, we choose
\begin{equation}
    g^I(x)=z_1^I Y_\ell(x)+z_2^IY_{\ell+1}(x)\,.
\end{equation}
With this choice, invariance of (\ref{eq:Hinvint}) reduces after some algebra to the recursive condition
\begin{equation}
(d+\ell-\Delta_I)H^{(\ell+1)}_{IJ}=(\ell+\Delta_I)H_{IJ}^{(\ell)}\,,
\end{equation}
which fixes the moments of $H_{IJ}|_{g=0}$ in terms of one normalisation $\mathcal{K}_{IJ}$,
\begin{equation}
    H^{(\ell)}_{IJ}=\frac{\Gamma(\ell+\Delta_I)}{\Gamma(d+\ell-\Delta_I)}\mathcal{K}_{IJ}\,.
\end{equation}
Since the spherical harmonics are complete, this normalisation fixes $H_{IJ}(x,y)|_{g=0}$ as a distribution, including its contact terms. It remains to identify this distribution. For $\Rea\Delta_I<\frac{d}{2}$ the function $s_{xy}^{-2\Delta_I}$ is integrable, so it defines a distribution. For other $\Delta_I$ we define the distribution $s_{xy}^{-2\Delta_I}$ by analytic continuation in $\Delta_I$. This continuation has poles where $2\Delta_I-d$ is a non-negative even integer, a case we discuss below. The quadratic form
\begin{equation}
    \frac{1}{2}\int \dd v(x)\,\dd v(y)\,g^I(x)C_{IJ}s_{xy}^{-2\Delta_I} g^J(y)
\end{equation}
is invariant under the same transformation (\ref{eq:gtransform}). For $\Rea\Delta_I<\frac{d}{2}$ this follows from the conformal covariance of $s_{xy}^{-2\Delta_I}$. Analytic continuation extends it to other $\Delta_I$. The moments of $s_{xy}^{-2\Delta_I}$, defined by
\begin{equation}
    \int_{S^d} \dd v(y)\, s_{xy}^{-2\Delta_I}Y_\ell(y)=s^{(\ell)}_{\Delta_I}Y_\ell(x)\,,
\end{equation}
therefore obey the same recursion relation,
\begin{equation}
    (d+\ell-\Delta_I)s^{(\ell+1)}_{\Delta_I}=(\ell+\Delta_I)s^{(\ell)}_{\Delta_I}\,.
\end{equation}
Hence $H_{IJ}(x,y)|_{g=0}$ is a constant multiple of the distribution $s_{xy}^{-2\Delta_I}$. Comparison with (\ref{eq:Hg0}) at non-coincident points fixes the constant to be $C_{IJ}$. Thus (\ref{eq:Hg0}) also holds at coincident points, as an identity between distributions.

This argument fails if $2\Delta_I-d=2k$ for a non-negative integer $k$, for which the ratio of Gamma functions in the recursion relation becomes a polynomial,
\begin{equation}
    \frac{\Gamma(\ell+\Delta_I)}{\Gamma(d+\ell-\Delta_I)}=\frac{\Gamma(\ell+d/2+k)}{\Gamma(\ell+d/2-k)}=\prod_{n=0}^{2k-1}\left(\ell+\frac{d}{2}-k+n\right)\,.
\end{equation}
The solution of the recursion is then a pure contact term, a polynomial in the Laplacian acting on $\delta(x,y)$, so it cannot be matched to the two-point function at non-coincident points. The two-point function then has no conformally covariant extension to coincident points, which signals an anomaly. However, since $d\notin 2\mathbb{N}$, $2\Delta_I-d$ is not a non-negative even integer for $\Delta_I$ close enough to $d$. Integrals of $H_{IJ}(x,y)|_{g=0}$ are then defined by analytic continuation of integrals of $s_{xy}^{-2\Delta_I}$. For instance, we can use the formula
\begin{equation}\label{eq:sxyintegral}
    \int_{S^d}\dd v(y)\,s_{xy}^{-2a}=\pi^{\frac{d}{2}}2^{d-2a}\frac{\Gamma\left(\frac{d}{2}-a\right)}{\Gamma(d-a)}\,,
\end{equation}
which holds for $\Rea a<\frac{d}{2}$, and for other $a$ provides the analytic continuation of the integral \cite{Cardy:1988cwa,Fei:2015oha}. Applying this formula to (\ref{eq:Gchordal}) gives
\begin{equation}
    \begin{split}
        G_{IJ}\big|_{g=0}&=\frac{2^{d-2\Delta_I+1}\pi^{\frac{d}{2}}V_d}{d}\frac{\Gamma\!\left(1+\frac{d}{2}-\Delta_I\right)\sin\frac{\pi d}{2}}{\Gamma(1+d-\Delta_I)}C_{IJ}\\
        &=\frac{2^{d-2\Delta_I+1}\pi^{\frac{d}{2}+1}V_d}{d\lsp\Gamma\!\left(\Delta_I-\frac{d}{2}\right)\Gamma(1+d-\Delta_I)}\frac{\sin\frac{\pi d}{2}}{\sin\left(\pi\left(\Delta_I-\frac{d}{2}\right)\right)}C_{IJ}\,,
    \end{split}
\end{equation}
where in the second line we have used $\Gamma(1-x)\Gamma(x)=\pi/\sin(\pi x)$ with $x=\Delta_I-\frac{d}{2}$. For $\frac{d}{2}<\Delta_I<d+1$ both gamma functions in the second line are positive, so that the sign of the coefficient of $C_{IJ}$ is the sign of the ratio of sines. This ratio equals one at $\Delta_I=d$ and changes sign only where $2\Delta_I-d$ is an even integer. The coefficient of $C_{IJ}$ is therefore positive whenever $2\Delta_I-d$ and $d$ lie between the same two consecutive even integers, which includes all $\Delta_I$ close to $d$. In particular, for $\Delta_I=d$ the metric is
\begin{equation}
    G_{IJ}|_{g=0}=\frac{2\pi^{d+1}}{\Gamma(d+1)}\,C_{IJ}\qquad (\Delta_I=d)\,.
\end{equation}

The perturbative positive-definiteness of $G_{IJ}$ thus hinges on the positive definiteness of the two-point function coefficient $C_{IJ}$. For integer $d$, this follows from unitarity of the CFT at $g^I=0$. In the case where the theory we are perturbing about is a free scalar theory, and the perturbing operators contain no derivatives, Wick's theorem permits a choice of basis in which $C_{IJ}=\delta_{IJ}$ up to a positive combinatorial factor in any $d$. In either case, $G_{IJ}$ is positive definite at weak coupling.

So far we have set the radius $r$ of the sphere to one. By dimensional analysis, $\ln Z$ depends on the renormalisation scale $\mu$ and on $r$ only through the product $\mu r$. Besides the couplings, the potential $A$ and the metric $G_{IJ}$ therefore depend on $\mu r$, which we hold fixed. Along a flow $g^I(\mu)$ we evaluate $A$ on the running couplings at this fixed value of $\mu r$, so that $\mu\,\dd A/\dd\mu=\bh^I\partial_IA$. Consider a flow that approaches a UV fixed point as $\mu\to\infty$ and an IR fixed point as $\mu\to0$, and that remains in the region where $G_{IJ}$ is positive definite. Integrating (\ref{eq:NAgrad}) along it gives a weak version of the $\widetilde{F}$-theorem,
\begin{equation}
    \widetilde{F}_{\text{UV}}-\widetilde{F}_{\text{IR}}=\int_0^\infty\dd\mu\,\frac{\dd A}{\dd\mu}=\int_0^\infty\frac{\dd\mu}{\mu}\,\Bh^IG_{IJ}\Bh^J\geq0\,,
\end{equation}
where we have used that $A$ is flavour invariant, so that $\bh^I\partial_IA=\Bh^I\partial_IA$. The value at which $\mu r$ is held fixed is arbitrary, because a change of $\mu r$ is equivalent to transport along the flow, as we show at the end of section \ref{sec:anom}.

The gradient condition (\ref{eq:NAgrad}) also holds for the long-range multiscalar $\phi^4$ theory in $0<d<4$ with $d\neq2$ \cite{Fisher:1972zz,Benedetti:2020rrq}. The kinetic term of this theory is a fractional power of the Laplacian. It is non-local, and the theory has no local energy-momentum tensor \cite{Paulos:2015lrconformal}, so it lies outside the assumptions of section \ref{sec:localRG}. The sphere Ward identity nevertheless holds for it, by a direct argument. The free theory is a generalised free field of dimension $\Delta_\phi$, whose propagator on the sphere is proportional to $s_{xy}^{-2\Delta_\phi}$. It is invariant under the conformal transformations of the sphere. The quartic operators have dimension $4\Delta_\phi$. The ultraviolet divergences are regulated analytically, by taking $\Delta_\phi$ slightly below $\frac d4$, which preserves this invariance. The bare partition function is therefore invariant when the bare couplings $g_0^I$ transform as the sources of operators of dimension $4\Delta_\phi$, that is under $\delta g_0^I=-(d-4\Delta_\phi)\lsp Yg_0^I+\partial^\mu Y\lsp\partial_\mu g_0^I$, as in (\ref{eq:gtransform}).

The divergences are local, so they do not renormalise the non-local kinetic term. Moreover, for $0<d<4$ with $d\neq2$ the only local terms of dimension $d$ allowed by parity and by the symmetry $\phi\to-\phi$ are quartic in the fields and contain no derivatives.\footnote{For $d=2$ there are also field-independent terms of dimension two, built from the curvature and from derivatives of the couplings, and the theory has an anomaly.} Renormalisation therefore only redefines the couplings, also when they depend on position, and the bare couplings are $\mu^{d-4\Delta_\phi}$ times functions of the renormalised couplings at the same point. Since the bare couplings do not depend on $\mu$, they obey $\bh^J\partial_Jg_0^I=-(d-4\Delta_\phi)\lsp g_0^I$. Their transformation is therefore that of the renormalised couplings under $\delta g^I=Y\bh^I+\partial^\mu Y\lsp\partial_\mu g^I$. This is the content of (\ref{eq:sphereWard}) with $\Bh^I=\bh^I$, so that (\ref{eq:NAgrad}) holds for the long-range theory to all orders.

\section{Theories with a Weyl anomaly}\label{sec:anom}
The presence of a Weyl anomaly spoils the argument of the previous section. With $\sigma=Y$, the local renormalisation group equation \eqref{eq:localRG} now reads $\Delta_Y\ln Z=\mathcal{A}_Y$, where $\mathcal{A}_Y$ is the anomaly associated with the transformation $\sigma(x)=Y(x)$. The diffeomorphism identity \eqref{eq:diffidentity} is unchanged, and subtracting it from \eqref{eq:localRG} we find that the sphere Ward identity (\ref{eq:sphereWard}) becomes
\begin{equation}\label{eq:sphereWardAnom}
    \int_{S^d} \dd^d x\left[(Y \Bh^I+\partial^\mu Y\lsp\partial_\mu g^I)\frac{\delta \ln Z(\gamma,g,a)}{\delta g^I}+(Y\tilde{\rho}_I D_\mu g^I-\stL_\xi\lsp a_\mu)\cdot\frac{\delta \ln Z(\gamma,g,a)}{\delta a_\mu}\right]=\mathcal{A}_Y\,.
\end{equation}
As before, we will only use this equation at $a_\mu=0$.

We work near an even dimension $D$, in $d=D-\varepsilon$ dimensions, using dimensional regularisation and taking account of Weyl anomalies. We keep $\varepsilon\neq0$ throughout and demand that all quantities be regular as $\varepsilon\to0$, so that we can set $\varepsilon=0$ at the end and recover $D$-dimensional results. Regularity requires counterterms with poles in $\varepsilon$, and these give rise to the anomaly. We have again omitted the sources of relevant operators, which we restore in the following section. Without them, the counterterms of dimensional regularisation are $\mu^{-\varepsilon}$ times integrals of local densities of dimension $D$, with poles in $\varepsilon$. We allow finite counterterms of the same form and exclude local terms with other powers of $\mu$, such as $\mu^{d-2}\int\dd v\,f(g)\lsp R$. Such a term would introduce a mass scale, and it does not arise in a mass-independent scheme. Every local term in $\ln Z$ is then $\mu^{-\varepsilon}$ times the integral of a local density of dimension $D$, and the same holds for the anomaly $\mathcal{A}_\sigma$.

The anomaly $\mathcal{A}_Y$ spoils the immediate use of the sphere Ward identity as a gradient condition. To see this, perturb the couplings by $z^IY(x)$. At constant couplings, rotational invariance makes $\mathcal{A}_Y$ proportional to $\int_{S^d}\dd v\,Y$, which vanishes by (\ref{eq:Yprops}). At first order in $z$, rotational invariance similarly makes $\mathcal{A}_Y$ proportional to $\int_{S^d}\dd v\,Y^2=N_Y$, and we may therefore write
\begin{equation}\label{eq:AnomOmegaDef}
    \mathcal{A}_Y=N_Yz^I\Omega_I+\text{O}(z^2)\,,
\end{equation}
which defines the one-form $\Omega_I$ on the space of couplings. The algebra of section \ref{sec:free} then gives
\begin{equation}\label{eq:SphereWardLinear}
    \partial_I\left(\frac{1}{V_d}(d+\Bh^J\partial_J)\ln Z\right)-(H^{(0)}-H^{(1)})_{IJ}\Bh^J=\Omega_I\,.
\end{equation}
For $\Omega_I=0$ this is (\ref{eq:NAgrad}) divided by $\frac{V_d}{d}\sin\frac{\pi d}{2}$. The one-form $\Omega_I$ spoils the gradient interpretation of (\ref{eq:SphereWardLinear}).

We now show that the response of this equation to a rescaling of the sphere removes the obstruction. For this purpose, we restore the radius $r$ of the sphere and keep $Y$ fixed as a function of the angles. Then
\begin{equation}
    \dd v,\, V_d,\,N_Y\sim r^d\,,\qquad s_{xy}^2\sim r^2\,,
\end{equation}
where $V_d$ now denotes the volume of the sphere of radius $r$. The conformal Killing vector becomes $\xi_\mu=-r^2\partial_\mu Y$, and (\ref{eq:SphereWardLinear}) holds at any $r$. By dimensional analysis, $\ln Z$ depends on $\mu$ and $r$ only through $\mu r$. We take derivatives with respect to $r$ at fixed $\mu$ and fixed couplings. On a sphere of radius $r$, $\mu^{-\varepsilon}$ times the integral of a local density of dimension $D$ is proportional to $(\mu r)^{-\varepsilon}$. This applies to $\mathcal{A}_Y$, so that (\ref{eq:AnomOmegaDef}) gives
\begin{equation}\label{eq:Omegascaling}
    (r\partial_r+d+\varepsilon)\Omega_I=0\,.
\end{equation}
Here locality of the anomaly is essential. The functions $\Bh^I$ do not depend on $r$, so $r\partial_r$ commutes with $\Bh^I$ and with $\partial_I$. Acting with $r\partial_r+d+\varepsilon$ on (\ref{eq:SphereWardLinear}) therefore gives
\begin{equation}
    \partial_I\left((r\partial_r+d+\varepsilon)\left(\frac{1}{V_d}(d+\Bh^J\partial_J)\ln Z\right)\right)-(r\partial_r+d+\varepsilon)(H^{(0)}-H^{(1)})_{IJ}\Bh^J=0\,.
\end{equation}
Since $V_d\sim r^d$, we have $(r\partial_r+d+\varepsilon)(f/V_d)=(r\partial_r+\varepsilon)f/V_d$ for any $f$. We further multiply by $\frac{V_d}{d\lsp\varepsilon}\sin\frac{\pi d}{2}$, and note that the factor of $1/\varepsilon$ does not make the result singular, since it is cancelled by the zero of
\begin{equation}\label{eq:sinexp}
    \sin\frac{\pi d}{2}=(-1)^{\frac{D}{2}+1}\frac{\pi\varepsilon}{2}+\text{O}(\varepsilon^3)\,.
\end{equation}
We thus obtain the gradient condition
\begin{equation}\label{eq:Agrad}
    \partial_IA=G_{IJ}\Bh^J\,,
\end{equation}
where the potential $A$ and the metric $G_{IJ}$ are
\begin{equation}\label{eq:AgradDef}
    \begin{gathered}
        A=\frac{\sin\frac{\pi d}{2}}{\varepsilon}\left(1+\frac{1}{d}\Bh^I\partial_I\right)(r\partial_r+\varepsilon)\ln Z\,,\\
        G_{IJ}=\frac{\sin\frac{\pi d}{2}}{d\lsp\varepsilon}(r\partial_r+\varepsilon)[V_d(H^{(0)}-H^{(1)})]_{IJ}\,.
    \end{gathered}
\end{equation}
Again, the symmetry of the metric is manifest.

To match the value of $A$ at a fixed point with $\widetilde{F}$, we must be careful about the convention used to define the free energy. For constant $\sigma$, a Weyl rescaling changes the radius of the sphere, and $-2\int\dd^d x\,\gamma^{\mu\nu}\delta/\delta\gamma^{\mu\nu}=r\partial_r$. At constant couplings and $a_\mu=0$, (\ref{eq:localRG}) with $\sigma=1$ thus becomes
\begin{equation}\label{eq:radiusRG}
    (r\partial_r+\Bh^I\partial_I)\ln Z=\mathcal{A}_1\,,
\end{equation}
where $\mathcal{A}_1$ is the anomaly at $\sigma=1$, which is proportional to $(\mu r)^{-\varepsilon}$. We write $\mathcal{A}_*$ for its value at a fixed point $g_*$, where $\Bh^I=0$. By (\ref{eq:1ptTO}), $r\partial_r\ln Z=\int_{S^d}\dd v\,\langle T^\mu{}_\mu\rangle$, so $\mathcal{A}_*$ is the integrated trace of the energy-momentum tensor at the fixed point. Because of the anomaly, $\ln Z|_{g_*}$ depends on the radius of the sphere,
\begin{equation}
    r\partial_r \ln Z|_{g_*}=\mathcal{A}_*=-\frac{1}{\varepsilon}r\partial_r \mathcal{A}_*\,,
\end{equation}
where in the last equality we have used that $\mathcal{A}_*\propto(\mu r)^{-\varepsilon}$. This implies that we can write
\begin{equation}\label{eq:ZtoZCFT}
    \ln Z|_{g_*}=\ln Z_{\text{CFT}}-\frac{\mathcal{A}_*}{\varepsilon}\,,
\end{equation}
where $\ln Z_{\text{CFT}}$ is independent of the radius and equal to minus the free energy computed in \cite{Fei:2015oha}. The radius-independent $\ln Z_{\text{CFT}}$ defines the value of $\widetilde{F}$ at fixed points,
\begin{equation}
    \widetilde{F}_*=\sin\frac{\pi d}{2}\,\ln Z_{\text{CFT}}\,.
\end{equation}
At a fixed point, the potential then reduces to
\begin{equation}
    A_*=\frac{\sin\frac{\pi d}{2}}{\varepsilon}(r\partial_r+\varepsilon)\ln Z|_{g_*}=\sin\frac{\pi d}{2}\,\ln Z_{\text{CFT}}=\widetilde{F}_*\,,
\end{equation}
completing the matching.

As in (\ref{eq:Gchordal}), we can write the metric as
\begin{equation}
    G_{IJ}=\frac{1}{2d\lsp\varepsilon}\sin\frac{\pi d}{2}\,(r\partial_r+\varepsilon)\left[\frac{1}{r^2}\int_{S^d}\dd v(x)\,\dd v(y)\,s_{xy}^2H_{IJ}(x,y)\right].
\end{equation}
Contact terms in $H_{IJ}$ now also come from the counterterms that remove poles in $\varepsilon$. These counterterms contain derivatives of the couplings, so the factor $s_{xy}^2$ does not remove their contact terms. However, these counterterms, and any finite ones, are $\mu^{-\varepsilon}$ times integrals of local densities of dimension $D$. Their contribution to the bracket is therefore proportional to $(\mu r)^{-\varepsilon}$, which $r\partial_r+\varepsilon$ annihilates. At $g^I=0$, what remains is the two-point function (\ref{eq:Hg0}). Since the couplings are dimensionless, it now carries a factor of $\mu^{2(d-\Delta_I)}$. In dimensional regularisation, its integrals are given by analytic continuation in $\Delta_I$. On a sphere of radius $r$, the right-hand side of (\ref{eq:sxyintegral}) acquires a factor of $r^{d-2a}$, and we then obtain
\begin{equation}
    G_{IJ}|_{g=0}=\left(2(d-\Delta_I)+\varepsilon\right)\frac{2^{d-2\Delta_I+1}\pi^\frac{d}{2} (\mu r)^{2(d-\Delta_I)}V_d}{d\lsp\varepsilon\lsp r^d}\frac{\Gamma\!\left(1+\frac{d}{2}-\Delta_I\right)\sin\frac{\pi d}{2}}{\Gamma(1+d-\Delta_I)}C_{IJ}\,.
\end{equation}
By assumption, $\Delta_I-d$ is of order $\varepsilon$. Using (\ref{eq:sinexp}), one finds that to leading order in $\varepsilon$ this expression simplifies to
\begin{equation}
    G_{IJ}|_{g=0}=\frac{2\pi^{D+1}}{D!}\,C_{IJ}+\text{O}(\varepsilon)\,.
\end{equation}
This is the value found in section \ref{sec:free} for $\Delta_I=d$, now at $d=D$. It does not depend on the ratio $(\Delta_I-d)/\varepsilon$.

The anomalous case thus proceeds in the same way as the non-anomalous one. $A$ matches $\widetilde{F}$ at all fixed points and, to leading order, $G_{IJ}$ equals the two-point function coefficient $C_{IJ}$ times a positive number. For unitary theories this guarantees that the metric is positive definite, so long as the higher-order terms in the perturbative expansion remain small. As in section \ref{sec:free}, we hold $\mu r$ fixed, so that $A$ and $G_{IJ}$ are functions of the couplings and $\mu\,\dd A/\dd\mu=\bh^I\partial_IA$ along a flow. For a flow between two fixed points that stays within such a region, we obtain a weak $\widetilde{F}$-theorem by integrating
\begin{equation}
    \widetilde{F}_{\text{UV}}-\widetilde{F}_{\text{IR}}=\int_0^\infty\dd\mu\lsp\frac{\dd A}{\dd \mu}=\int_0^\infty\frac{\dd\mu}{\mu}\,\Bh^IG_{IJ}\Bh^J\geq0\,,
\end{equation}
where we have again used that $A$ is flavour invariant. All of these objects are regular in $\varepsilon$ by construction, so the limit $\varepsilon\to0$ is smooth. In this limit, the potential at fixed points becomes proportional to the integrated anomaly on the sphere. Indeed, $\ln Z|_{g_*}$ is regular, so $\sin\frac{\pi d}{2}\,\ln Z|_{g_*}$ vanishes as $\varepsilon\to0$. Writing $\ln Z_{\text{CFT}}=\ln Z|_{g_*}+\mathcal{A}_*/\varepsilon$ and using (\ref{eq:sinexp}), we find
\begin{equation}
    A_*=\frac{(-1)^{\frac{D}{2}+1}\pi}{2}\mathcal{A}_*\,,
\end{equation}
which is consistent with Cardy's monotonicity proposal~\cite{Cardy:1988cwa}. In $D=2$ we have $\mathcal{A}_*=c/3$, so that $A_*=\pi c/6$, where $c$ is equivalently the CFT's central charge and the coefficient of the Euler anomaly. In $D=4$ we have $\mathcal{A}_*=-\sA$ in the conventions of section \ref{sec:fourd}, so that $A_*=\pi\sA/2$, where $\sA$ is the coefficient of the Euler anomaly. Both values agree with $\widetilde{F}$ as computed in \cite{Giombi:2014xxa}.

The potential and the metric depend on the dimensionless product $\mu r$. We have kept it fixed, but its value is arbitrary. Indeed, a change of $\mu r$ is equivalent to transport along the flow. To see this, we note that $r\partial_r+\varepsilon$ annihilates $\mathcal{A}_1$ and commutes with $r\partial_r+\Bh^I\partial_I$. Acting with it on (\ref{eq:radiusRG}) gives
\begin{equation}
    (r\partial_r+\Bh^I\partial_I)(r\partial_r+\varepsilon)\ln Z=0\,.
\end{equation}
The operator $\Bh^I\partial_I$ also commutes with $r\partial_r+\Bh^J\partial_J$. By (\ref{eq:AgradDef}), the potential therefore satisfies
\begin{equation}
    (r\partial_r+\bh^I\partial_I)A=0\,,
\end{equation}
where we have used that $A$ is flavour invariant. Increasing the radius thus has the same effect on $A$ as lowering the renormalisation scale by the same factor. The same holds for $\Bh^IG_{IJ}\Bh^J=\Bh^I\partial_IA$, the integrand of the $\widetilde{F}$-theorem. In particular, $A$ does not change when $\mu$ and the couplings vary together along the flow at a fixed radius. The $\widetilde{F}$-theorem refers instead to the flow of the couplings at fixed $\mu r$. Equivalently, $r\partial_rA=-\Bh^IG_{IJ}\Bh^J\leq0$ at fixed $\mu$ and fixed couplings, so that $A$ decreases as the sphere grows. The potentials obtained for different values of $\mu r$ are therefore related by the flow, and all of them equal $\widetilde{F}_*$ at fixed points. Without an anomaly, the same relations follow from (\ref{eq:radiusRG}) with $\mathcal{A}_1=0$, for the potential of (\ref{eq:NAgrad}).

\section{The effect of relevant sources}\label{sec:relsource}
In this section we restore the sources of relevant operators, which we have set to zero so far. On the sphere this is not consistent, because these sources mix with the curvature and with derivatives of the couplings. Sources of this type add a term to the sphere Ward identity, which could spoil the gradient condition. Section \ref{sec:relsource-ward} sets up the Ward identity with sources. Section \ref{sec:relsource-elimination} shows that the additional term drops out when the sources are fixed by a certain stationarity condition, and derives the resulting gradient condition. Without an anomaly, the stationarity condition states that the one-point functions of the relevant operators vanish on the sphere. Section \ref{sec:relsource-fixed} shows that the potential still equals $\widetilde{F}$ at fixed points and that the leading term in the metric is unchanged. The construction requires an invertible matrix of second derivatives with respect to the sources, and section \ref{sec:relsource-invert} specifies when this is the case.

\subsection{The Ward identity with sources}\label{sec:relsource-ward}
We work near an integer dimension $D$, with $d=D-\varepsilon$. Let $N^A$ be scalar sources of relevant operators. At $\varepsilon=0$ they have dimension $w_A$, with $0<w_A<D$. The index $A$ includes flavour components. An example is given by the mass sources of dimension two in four dimensions \cite{Jack:2013sha}. The curvature, pairs of derivatives, the field strength of $a_\mu$ and sources of even dimension form only scalars of even dimension, while a contraction with the Levi--Civita tensor has dimension at least $D$. The Weyl variation of a source whose dimension at $\varepsilon=0$ is not an even integer therefore vanishes when all such sources vanish. We set these sources to zero and keep only sources of even dimension.

The operator $\Delta_\sigma$ of (\ref{eq:localRG}) gains the term $\int\dd^d x\,(\Delta_\sigma N^A)\lsp\delta/\delta N^A$, and the Weyl variation $\Delta_\sigma N^A$ is a local expression of dimension $w_A$, linear in $\sigma$. Besides the scaling term $-\sigma\lsp w_AN^A$, it contains the anomalous dimensions of the sources and mixing terms built from the curvature, derivatives of the couplings and sources of lower dimension. It can also contain derivatives of $\sigma$. For the mass sources near four dimensions, for example, the mixing terms are $R$, $D^2g^I$ and $D_\mu g^ID^\mu g^J$, with coefficients that depend on the couplings \cite{Jack:2013sha}. In this section we make no assumption on the form of $\Delta_\sigma N^A$. We multiply each source by a power of $\mu$, so that its dimension is exactly $w_A$ for every $\varepsilon$ and the dimensional analysis below then involves only integer dimensions. The functions $\Bh^I$ and $\tilde\rho_I$ are dimensionless, so they do not depend on $N^A$.

Let $\mathcal{D}$ be the operator that rescales the metric by a constant factor and each source according to its dimension,
\begin{equation}\label{eq:scaling-operator}
    \mathcal{D}=\int\dd^d x\left[-2\gamma^{\mu\nu}\frac{\delta}{\delta\gamma^{\mu\nu}}-\sum_Aw_AN^A\frac{\delta}{\delta N^A}\right]\,.
\end{equation}
It leaves the couplings and $a_\mu$ unchanged. The coefficients in $\Delta_\sigma$ are dimensionless functions of the couplings. Under this rescaling, the Weyl variations of the metric, the couplings, $a_\mu$ and the sources therefore scale in the same way as these quantities themselves, so that $[\mathcal{D},\Delta_\sigma]=0$. When there is an anomaly, we assume as in section \ref{sec:anom} that it is $\mu^{-\varepsilon}$ times the integral of a local density of dimension $D$, now built also from the sources. This implies that $(\mathcal{D}+\varepsilon)\lsp\mathcal{A}_\sigma=0$. We define
\begin{equation}\label{eq:U-def}
    U=\sin\frac{\pi d}{2}\,\ln Z\qquad\text{or}\qquad U=\frac{\sin\frac{\pi d}{2}}{\varepsilon}\lsp(\mathcal{D}+\varepsilon)\ln Z\,,
\end{equation}
without and with an anomaly, respectively. In both cases $U$ is regular and obeys the local renormalisation group equation without an anomaly,
\begin{equation}\label{eq:U-localRG}
    \Delta_\sigma U=0\,.
\end{equation}
It is also invariant under diffeomorphisms and flavour transformations, since $\mathcal{D}$ commutes with them. Without sources $\mathcal{D}$ acts on the sphere as $r\partial_r$, and $(1+\frac1d\Bh^I\partial_I)\lsp U$ is the potential of (\ref{eq:NAgrad}) or (\ref{eq:AgradDef}).

The derivation of section \ref{sec:free} therefore applies to $U$, on a sphere of any radius $r$. With the conformal Killing vector $\xi_\mu=-r^2\partial_\mu Y$ of section \ref{sec:anom}, it gives
\begin{equation}\label{eq:sphereWardU}
    \int_{S^d}\dd^d x\left[(Y\Bh^I-\xi^\mu\partial_\mu g^I)\frac{\delta U}{\delta g^I}+(\Delta_YN^A-\xi^\mu\partial_\mu N^A)\frac{\delta U}{\delta N^A}+Y\tilde\rho_I\lsp\partial_\mu g^I\cdot\frac{\delta U}{\delta a_\mu}\right]=0
\end{equation}
at $a_\mu=0$, for any couplings $g^I(x)$ and sources $N^A(x)$.

\subsection{Elimination of the sources}\label{sec:relsource-elimination}
At constant couplings and constant sources $N^A=n^A$ on the sphere, $U$ is a function $U(g,n)$. It also depends on $\mu r$, which we keep fixed as in section \ref{sec:anom}. We write $\partial_A=\partial/\partial n^A$. In analogy with (\ref{eq:HIJdefn}) and (\ref{eq:angmoms}), we define the angular moments of the second derivatives of $U$ at this configuration,
\begin{equation}\label{eq:U-moments}
    U_{IJ}(x,y)=\frac{1}{\sqrt{\gamma(x)}\sqrt{\gamma(y)}}\frac{\delta^2U}{\delta g^I(x)\,\delta g^J(y)}\,,\qquad\int_{S^d}\dd v(y)\,U_{IJ}(x,y)\lsp Y_\ell(y)=U^{(\ell)}_{IJ}\lsp Y_\ell(x)\,,
\end{equation}
and in the same way $U^{(\ell)}_{IA}=U^{(\ell)}_{AI}$ and $U^{(\ell)}_{AB}$, with derivatives with respect to the corresponding sources. The zeroth moments are the second derivatives of the function $U(g,n)$ divided by $V_d$, so that $V_dU^{(0)}_{AB}=\partial_A\partial_BU$, for example. We assume for now that the matrices $U^{(0)}_{AB}$ and $U^{(1)}_{AB}$ have regular inverses, which we denote by $U_{(0)}^{AB}$ and $U_{(1)}^{AB}$. Section \ref{sec:relsource-invert} discusses in detail when this is the case.

We fix the sources in two steps. First, at constant couplings we choose the constant sources $n^A=n_0^A(g)$ that solve
\begin{equation}\label{eq:stationary}
    \partial_AU(g,n)=0\,.
\end{equation}
Since $U^{(0)}_{AB}$ is invertible, the solution is unique in perturbation theory and regular. We call it the stationary configuration. By rotational invariance, $\delta U/\delta N^A$ is constant on the sphere at this configuration, so (\ref{eq:stationary}) states that it vanishes. Without an anomaly, this means that the one-point functions of the relevant operators vanish. With an anomaly, the same holds after the action of $\mathcal{D}+\varepsilon$, which removes local terms. We define
\begin{equation}\label{eq:Ubar}
    \bar U(g)=U(g,n_0(g))\,,
\end{equation}
which is flavour invariant. Second, we perturb the couplings as before, $g^I(x)=g^I+z^IY(x)$, and let the sources follow them,
\begin{equation}\label{eq:source-response}
    N^A(x)=n_0^A(g)+\zeta^A{}_Iz^I\lsp Y(x)\,,\qquad\zeta^A{}_I=-U_{(1)}^{AB}\lsp U^{(1)}_{BI}\,.
\end{equation}
At this configuration the first derivatives of $U$ are
\begin{equation}\label{eq:U-first-order}
    \frac{1}{\sqrt\gamma}\frac{\delta U}{\delta N^A}=\text{O}(z^2)\,,\qquad\frac{1}{\sqrt\gamma}\frac{\delta U}{\delta g^I}=\frac{1}{V_d}\lsp\partial_I\bar U+\bar U^{(1)}_{IJ}\lsp z^J\lsp Y+\text{O}(z^2)\,,
\end{equation}
where
\begin{equation}\label{eq:Ubar-moments}
    \bar U^{(\ell)}_{IJ}=U^{(\ell)}_{IJ}-U^{(\ell)}_{IA}\lsp U_{(\ell)}^{AB}\lsp U^{(\ell)}_{BJ}\,.
\end{equation}
Indeed, the first-order term of $\delta U/\delta N^A$ is proportional to $(U^{(1)}_{AI}+U^{(1)}_{AB}\lsp\zeta^B{}_I)\lsp z^I$, which vanishes by the choice of $\zeta^A{}_I$. The zeroth-order term of $\delta U/\delta g^I$ is the derivative of $U(g,n)$ with respect to $g^I$ at fixed $n^A$, divided by $V_d$. By (\ref{eq:stationary}) it equals $\partial_I\bar U/V_d$. Differentiating (\ref{eq:stationary}) gives $\partial_In_0^A=-U_{(0)}^{AB}\lsp U^{(0)}_{BI}$, and therefore
\begin{equation}\label{eq:Ubar-hessian}
    \partial_I\partial_J\bar U=V_d\lsp\bar U^{(0)}_{IJ}\,.
\end{equation}

We now evaluate the Ward identity (\ref{eq:sphereWardU}) to first order in $z$. The term with $\delta U/\delta N^A$ is of order $z^2$, whatever the form of $\Delta_YN^A$. The term with $\delta U/\delta a_\mu$ is of order $z^2$ as well, because $\delta U/\delta a_\mu$ vanishes at constant couplings and sources by rotational invariance. The remaining term is that of section \ref{sec:free}, with $\ln Z$ and $H^{(1)}_{IJ}$ replaced by $\bar U$ and $\bar U^{(1)}_{IJ}$, so that (\ref{eq:linWard}) becomes
\begin{equation}\label{eq:linWardU}
    (d\lsp\delta_I{}^J+\partial_I\Bh^J)\,\partial_J\bar U+V_d\lsp\bar U^{(1)}_{IJ}\Bh^J=0\,.
\end{equation}
With (\ref{eq:Ubar-hessian}), this is the gradient condition
\begin{equation}\label{eq:gradient-sources}
    \partial_IA=G_{IJ}\Bh^J\,,\qquad A=\left(1+\frac{1}{d}\Bh^I\partial_I\lnsp\right)\bar U\,,\qquad G_{IJ}=\frac{V_d}{d}\lsp(\bar U^{(0)}-\bar U^{(1)})_{IJ}\,.
\end{equation}
The metric is manifestly symmetric. Without sources, $\bar U=U$ and these expressions reduce to (\ref{eq:NAgrad}) and (\ref{eq:AgradDef}), depending on the choice at \eqref{eq:U-def}.

The potential and the metric do not change under a redefinition of the sources, $N^A\to N^A-\Xi^A$, with $\Xi^A$ a local expression of dimension $w_A$ built from the curvature, the couplings and sources of lower dimension. At fixed couplings, such a redefinition is a change of variables for the constant sources $n^A$ and for their first-harmonic perturbations. It is invertible, because $\Xi^A$ contains only sources of lower dimension. It maps the solution of (\ref{eq:stationary}) to the solution of the same equation in the new variables, so that $\bar U$ does not change. At the stationary configuration the first derivatives of $U$ with respect to the sources vanish, so that its second derivatives transform linearly. In each of the two harmonic sectors, the perturbations of the sources are replaced by invertible linear combinations of themselves, shifted by multiples of the perturbations of the couplings. The combinations (\ref{eq:Ubar-moments}) are invariant under such a change. Redefinitions of this kind change the form of $\Delta_\sigma N^A$, and in particular its terms with derivatives of $\sigma$. This is why the form of $\Delta_\sigma N^A$ plays no role.

As in section \ref{sec:anom}, a change of $\mu r$ is equivalent to transport along the flow. At constant couplings and sources, (\ref{eq:U-localRG}) with $\sigma=1$ reads $(r\partial_r+\Bh^I\partial_I+\Delta_1N^A\lsp\partial_A)\lsp U=0$, where $r\partial_r$ and $\partial_I$ act at fixed $n^A$. At $n^A=n_0^A(g)$ the last term vanishes by (\ref{eq:stationary}), and for the same reason the other two derivatives of $U$ equal those of $\bar U$. Hence
\begin{equation}\label{eq:Ubar-radius}
    (r\partial_r+\Bh^I\partial_I)\lsp\bar U=0\,,
\end{equation}
and $(r\partial_r+\bh^I\partial_I)A=0$ follows as in section \ref{sec:anom}.

\subsection{Fixed points and the leading metric}\label{sec:relsource-fixed}
We first show that $A$ still equals $\widetilde{F}$ at fixed points. At constant couplings and sources, every term of $\Delta_YN^A$ is proportional to $Y$, because $\nabla_\mu\partial_\nu Y=-Y\gamma_{\mu\nu}/r^2$. We write $\Delta_YN^A=-Y\lsp \vartheta^A(g,n)$. We vary the Ward identity (\ref{eq:sphereWardU}) with respect to the sources by $\delta N^A=z^AY$, with constant $z^A$, at constant couplings and at $n^A=n_0^A(g)$. Since $\delta U/\delta N^A$ vanishes at this configuration, the second term of (\ref{eq:sphereWardU}) contributes only through the variation of $\delta U/\delta N^A$, and we find
\begin{equation}\label{eq:v-relation}
    U^{(1)}_{AB}\lsp \vartheta^B=U^{(1)}_{AI}\Bh^I\,.
\end{equation}
Thus $\vartheta^A=-\zeta^A{}_I\Bh^I$. On the stationary configuration, the Weyl variation of the sources is therefore the change (\ref{eq:source-response}) induced by the Weyl variation $Y\Bh^I$ of the couplings.

At a fixed point $g_*$, (\ref{eq:v-relation}) gives $\vartheta^A=0$, so that $\Delta_YN^A=0$. Since also $\Delta_Yg^I=Y\Bh^I=0$, the configuration is invariant under the conformal transformations of the sphere. The theory on the sphere with the sources $n_0^A(g_*)$ is therefore the fixed-point CFT. Without an anomaly, $\bar U(g_*)=\sin\frac{\pi d}{2}\,\ln Z$ is then $\widetilde{F}_*$. With an anomaly we follow section \ref{sec:anom}. The condition $\vartheta^A=0$ has a unique solution in perturbation theory, because at $g^I=0$ and $\varepsilon=0$ the matrix $\partial\vartheta^A/\partial n^B$ equals $w_A\lsp\delta^A{}_B$ up to terms with $w_B<w_A$. The condition contains no scale other than $r$, so its solution is $n_0^A(g_*)\propto r^{-w_A}$ at fixed $\mu$, and $\mathcal{D}$ acts on functions of this configuration as $r\lsp\dd/\dd r$. By (\ref{eq:Ubar-radius}), $\bar U(g_*)$ does not depend on $r$. The definition (\ref{eq:U-def}) then gives
\begin{equation}\label{eq:lnZ-fixed-sources}
    \left(r\frac{\dd}{\dd r}+\varepsilon\right)\ln Z=\frac{\varepsilon\lsp\bar U(g_*)}{\sin\frac{\pi d}{2}}\,,
\end{equation}
so that $\ln Z$ is the sum of a constant and a multiple of $(\mu r)^{-\varepsilon}$. As in (\ref{eq:ZtoZCFT}), the constant is the radius-independent $\ln Z_{\text{CFT}}$, and so
\begin{equation}
    A_*=\bar U(g_*)=\sin\frac{\pi d}{2}\,\ln Z_{\text{CFT}}=\widetilde{F}_*\,.
\end{equation}

Next we show that the leading term in the metric is unchanged. At $g^I=0$ the theory is the reference CFT. We assume that its relevant operators can be taken to be primaries, as in a unitary CFT. The potential and the metric do not change under redefinitions of the sources, so we may take the sources to couple to these primaries. At linear order the couplings and the sources then transform as in (\ref{eq:gtransform}), each with the dimension of its operator. As in section \ref{sec:free}, the Ward identity forbids linear terms in the expansion of $U$ around $g^I=0$ and $N^A=0$, so that $n_0^A=0$ at $g^I=0$, and it constrains the quadratic terms. For two operators of dimensions $\Delta$ and $\Delta'$, the moments of the corresponding second derivative of $U$ obey
\begin{equation}\label{eq:mixed-recursion}
    (d+\ell-\Delta)\lsp U^{(\ell+1)}=(\ell+\Delta')\lsp U^{(\ell)}\,,\qquad(d+\ell-\Delta')\lsp U^{(\ell+1)}=(\ell+\Delta)\lsp U^{(\ell)}\,,
\end{equation}
which for $\Delta=\Delta'$ is the recursion of section \ref{sec:free}. For $\Delta\neq\Delta'$ and $\Delta+\Delta'\neq d$ the two relations are compatible only if all the moments vanish. Hence $U^{(\ell)}_{IA}$ vanishes at $g^I=0$. So does the difference $\bar U^{(\ell)}_{IJ}-U^{(\ell)}_{IJ}$, by (\ref{eq:Ubar-moments}). The leading term of $G_{IJ}$ is therefore that of sections \ref{sec:free} and \ref{sec:anom}.

\subsection{When the sources can be eliminated}\label{sec:relsource-invert}
It remains to determine when $U^{(0)}_{AB}$ and $U^{(1)}_{AB}$ have regular inverses. This is the case if they are invertible at leading order, that is at $g^I=0$ and $\varepsilon=0$. As in section \ref{sec:relsource-fixed}, we take the sources to couple to primaries of the reference CFT. At $g^I=0$ and $N^A=0$, the second derivative of $\ln Z$ with respect to the sources is then the two-point function of these operators, up to local terms,
\begin{equation}\label{eq:source-twopoint}
    \frac{1}{\sqrt{\gamma(x)}\sqrt{\gamma(y)}}\frac{\delta^2\ln Z}{\delta N^A(x)\,\delta N^B(y)}=\frac{\mu^{\varepsilon_A-\varepsilon}\lsp C_{AB}}{(s_{xy}^2)^{\Delta_A}}\,,\qquad\varepsilon_A=2(d-w_A-\Delta_A)+\varepsilon\,.
\end{equation}
Here $\Delta_A$ is the dimension of the operator sourced by $N^A$, and the matrix $C_{AB}$ of two-point function coefficients vanishes between operators of different dimensions. The power of $\mu$ appears because the sources have dimension exactly $w_A$, and $\varepsilon_A$ is of order $\varepsilon$. The moments of the right-hand side follow from (\ref{eq:sxyintegral}) and the recursion relation of section \ref{sec:free}. On the unit sphere,
\begin{equation}\label{eq:s-moments}
    \int_{S^d}\dd v(y)\,s_{xy}^{-2\Delta}\lsp Y_\ell(y)=\pi^{\frac d2}\lsp2^{d-2\Delta}\frac{\Gamma\!\left(\frac d2-\Delta\right)\Gamma(\ell+\Delta)}{\Gamma(\Delta)\lsp\Gamma(d+\ell-\Delta)}\lsp Y_\ell(x)\,.
\end{equation}

Near odd $D$ there is no anomaly, and no local term of dimension $D$ can be built with sources of even dimension. At $g^I=0$, $U^{(\ell)}_{AB}$ is then $\mu^{\varepsilon_A-\varepsilon}\lsp C_{AB}\sin\frac{\pi d}{2}$ times the coefficient in (\ref{eq:s-moments}) with $\Delta=\Delta_A$. At $d=D$ and $\Delta_A=D-w_A$, with $D$ odd and $0<w_A<D$, the argument of the first gamma function is a half-integer and the arguments of the other three are positive integers. The coefficient is therefore finite and non-zero for every $\ell$.

Near even $D$ the local terms are annihilated by $\mathcal{D}+\varepsilon$, as in section \ref{sec:anom}, while the term of $\ln Z$ that is quadratic in the sources, with the kernel (\ref{eq:source-twopoint}), is an eigenfunction of this operator with the eigenvalue $\varepsilon_A$. On the unit sphere we therefore find
\begin{equation}\label{eq:U-NN-even}
    U^{(\ell)}_{AB}\big|_{g=0}=\frac{\varepsilon_A}{\varepsilon}\lsp\pi^{\frac d2}\lsp2^{d-2\Delta_A}\mu^{\varepsilon_A-\varepsilon}\frac{\Gamma\!\left(\frac d2-\Delta_A\right)\Gamma(\ell+\Delta_A)\sin\frac{\pi d}{2}}{\Gamma(\Delta_A)\lsp\Gamma(d+\ell-\Delta_A)}\lsp C_{AB}\,.
\end{equation}
For $w_A\leq D/2$ the gamma function $\Gamma(\frac d2-\Delta_A)=\Gamma(w_A-\frac D2+\frac{\varepsilon_A}{2})$ has a pole at $\varepsilon_A=0$, which the prefactor $\varepsilon_A$ cancels. With (\ref{eq:sinexp}) and even $w_A$, we obtain
\begin{equation}\label{eq:U-NN-leading}
    U^{(\ell)}_{AB}\big|_{g=0}=-\frac{\pi^{\frac D2+1}\lsp2^{2w_A-D}}{\left(\frac D2-w_A\right)!}\frac{\Gamma(\ell+D-w_A)}{\Gamma(D-w_A)\lsp\Gamma(\ell+w_A)}\lsp C_{AB}+\text{O}(\varepsilon)\,.
\end{equation}
The leading term does not depend on the ratio $\varepsilon_A/\varepsilon$, and it is non-zero for every $\ell$. For the mass sources near four dimensions, for example, it equals $-\pi^3C_{AB}$ for every $\ell$. Its sign plays no role, since only the inverse is needed. For $w_A>D/2$ there is no pole, and in that case $U^{(\ell)}_{AB}$ is of order $\varepsilon$ at $g^I=0$ and its inverse is not regular.

In a unitary theory $C_{AB}$ is positive definite. The sources can therefore be eliminated near odd dimensions, and near an even dimension $D$ if every source that we keep has $w_A\leq D/2$. This includes the mass sources near four dimensions. The condition is restrictive in one case only. In a unitary theory a scalar operator other than the identity has dimension at least $\frac{D-2}{2}$, so that $w_A\leq\frac D2+1$. Near even $D$ the condition can thus fail only for $w_A=\frac D2+1$, the source of a free scalar field. This source is kept only if its dimension is even, that is for $D=6,10,\ldots\,$. In four dimensions its dimension is three, and near three dimensions it is a half-integer. In two dimensions it would have $w_A=D$, a case that section \ref{sec:twod} excludes.

Among the theories considered here, only $\phi^3$ theory near six dimensions is affected. The source $N$ of $\phi$ has $w_N=4$ and $\varepsilon_N=0$ in the free theory. There, $\ln Z$ depends on $N$ only through a term quadratic in $N$, on which $\mathcal{D}$ has the eigenvalue $-\varepsilon$. This term is annihilated by $\mathcal{D}+\varepsilon$, so that $U$ does not depend on $N$ at $g^I=0$. The inverse of $U^{(\ell)}_{NN}$ is then singular at weak coupling, and the construction of this section does not apply. The source cannot simply be set to zero either, because its Weyl variation contains terms without $N$. Therefore, whether the gradient condition holds to all loop orders for $\phi^3$ theory near six dimensions remains open.

\section{Gradient flow in particular dimensions}\label{sec:specificd}
We now compare the general construction with the local renormalisation group in two, four and six dimensions. As noted in the introduction, the Weyl consistency conditions in these dimensions give a gradient-like equation with a one-form obstruction. In two and four dimensions, we show that the sphere Ward identity supplies an additional condition on the anomaly coefficients, which allows this one-form to be removed by finite counterterms. In six dimensions, we compare our construction with the published consistency conditions.

We call a one-form $e_I$ on the space of couplings removable if
\begin{equation}\label{eq:removable}
    e_I=\partial_If+h_{IJ}\Bh^J\,,
\end{equation}
with $f$ regular and flavour invariant and $h_{IJ}$ regular and symmetric. If $\partial_IA=G_{IJ}\Bh^J+e_I$ with a removable $e_I$, then $\partial_I(A-f)=(G+h)_{IJ}\Bh^J$, which is again a gradient condition with a symmetric metric.

\subsection{Two dimensions}\label{sec:twod}
Following the two-dimensional local renormalisation group of \cite{Osborn:1991gm,Jack:2013sha}, we write the parity-even anomaly as
\begin{equation}
    \mathcal{A}_\sigma=\frac{1}{2\pi}\int\dd v\,\big[\sigma(\sC\lsp R-\sG_{IJ}\lsp D_\mu g^ID^\mu g^J)-2\lsp \partial_\mu\sigma\lsp\sw_I\lsp D^\mu g^I\big]\,.
    \label{eq:twod_anomaly}
\end{equation}
This is (3.1) of \cite{Jack:2013sha}, with their $W$ replaced by $\ln Z$ and $\partial_\mu g^I$ by $D_\mu g^I$. The coefficients $\sC$, $\sG_{IJ}$ and $\sw_I$ are functions of the couplings. For a CFT, $\sC=c/12$, where $c$ is the central charge. The integrated anomaly of section \ref{sec:anom} is then $\mathcal{A}_*=4\sC=c/3$.

The operators $\Delta_\sigma$ commute, so that $\Delta_\sigma\mathcal{A}_{\sigma'}=\Delta_{\sigma'}\mathcal{A}_\sigma$. For (\ref{eq:twod_anomaly}) this Weyl consistency condition reads
\begin{equation}
    \partial_I\sC=\sG_{IJ}\Bh^J-(\stL_{\Bh,\tilde\rho}\lsp\sw)_I\,,
    \label{eq:twod_consistency}
\end{equation}
where $\stL_{\Bh,\tilde\rho}$ is the extended Lie derivative of \cite{Jack:2013sha}. It acts on one-forms and on symmetric tensors as
\begin{equation}
    (\stL_{\Bh,\tilde\rho}\lsp e )_I=(\stL_{\Bh}e)_I+\mathcal R_I{}^Je_J\,,\qquad (\stL_{\Bh,\tilde\rho}\lsp h)_{IJ}=(\stL_{\Bh}h)_{IJ}+\mathcal R_I{}^K h_{KJ}+\mathcal R_J{}^K h_{IK}\,,
    \label{eq:extended_lie}
\end{equation}
where $\stL_{\Bh}$ is the ordinary Lie derivative along $\Bh^I$ and $\mathcal R_I{}^J=(\tilde\rho_Ig)^J$. The terms with $\mathcal R_I{}^J$ arise from the variation of $a_\mu$ in (\ref{eq:localRG}). Using $(\stL_{\Bh}\sw)_I=\partial_I(\sw_J\Bh^J)-(\partial_I\sw_J-\partial_J\sw_I)\Bh^J$, the consistency condition becomes
\begin{equation}
    \partial_I\widetilde\sC=(\sG_{IJ}+\partial_I\sw_J-\partial_J\sw_I)\Bh^J-\mathcal R_I{}^J\sw_J\,,\qquad \widetilde\sC=\sC+\sw_I\Bh^I\,.
    \label{eq:twod_obstruction}
\end{equation}
This equation suffices to demonstrate strong monotonicity in two dimensions. Contracting with $\Bh^I$ projects out the antisymmetric part of the metric and also removes the term $\mathcal R_I{}^J\sw_J$,\footnote{This follows from the identity $\tilde\rho_I\Bh^I=0$, a consequence of the commutativity of Weyl rescalings on $a_\mu$ \cite{Jack:2013sha}.} so that
\begin{equation}
    \mu\frac{\dd\widetilde\sC}{\dd\mu}=\Bh^I\sG_{IJ}\Bh^J\geq0\,,
\end{equation}
where we have used that $\widetilde\sC$ is flavour invariant. Positivity is guaranteed perturbatively, because $\sG_{IJ}=\frac{\pi^2}{2}C_{IJ}+\text{O}(g)$ is proportional to the Zamolodchikov metric at leading order, which is positive by reflection positivity. At fixed points we have
\begin{equation}
    \widetilde\sC|_{g_*}=\frac{c_*}{12}\,,
\end{equation}
so that the monotone matches the central charge, in accordance with Zamolodchikov's $c$-theorem~\cite{Zamolodchikov:1986gt}. Nevertheless, the antisymmetric part $\partial_I\sw_J-\partial_J\sw_I$ of the metric and the extra term $\mathcal R_I{}^J\sw_J$ spoil any gradient conclusion that relies solely on the Weyl consistency conditions. The sphere Ward identity produces the gradient condition (\ref{eq:Agrad}), whose potential matches $\widetilde\sC$ at fixed points up to an overall positive factor. We therefore ask how this identity adds to the Weyl consistency conditions and resolves the difficulty with $\sw_I$.

The difficulty is smaller than it first appears, because finite local counterterms shift $\sw_I$. If we add the counterterms
\begin{equation}
    \ln Z\to\ln Z+\frac{1}{2\pi}\int\dd v\,\big(b\lsp R+c_{IJ}\lsp D_\mu g^I D^\mu g^J\big)\,,\qquad c_{IJ}=c_{JI}\,,
    \label{eq:twod_counterterm}
\end{equation}
where $b$ and $c_{IJ}$ are regular functions of the couplings, they shift the anomaly $\mathcal{A}_\sigma$ by
\begin{equation}
    \frac{1}{2\pi}\int\dd v\,\Big[\sigma\Big(R\lsp\Bh^J\partial_J b+(\stL_{\Bh,\tilde\rho}\lsp c)_{IJ}\lsp D_\mu g^I D^\mu g^J\Big)+2\lsp\partial_\mu\sigma\lsp\big(\partial_Ib+c_{IJ}\Bh^J\big) D^\mu g^I\Big]\,.
    \label{eq:twod_counterterm_variation}
\end{equation}
This corresponds to a shift of the anomaly coefficients by
\begin{equation}
    \sC\to\sC+\Bh^I\partial_I b\,,\qquad\sw_I\to\sw_I-\partial_Ib-c_{IJ}\Bh^J\,,\qquad\sG_{IJ}\to\sG_{IJ}-(\stL_{\Bh,\tilde{\rho}}\lsp c)_{IJ}\,.
    \label{eq:twod_scheme_shift}
\end{equation}
If $\sw_I$ is removable in the sense of (\ref{eq:removable}), so that $\sw_I=\partial_If+h_{IJ}\Bh^J$, we can set $b=f$ and $c_{IJ}=h_{IJ}$ to remove $\sw_I$ entirely. This choice also shifts $\sC$ and $\sG_{IJ}$, and the condition (\ref{eq:twod_obstruction}) becomes
\begin{equation}
    \partial_I(\sC+\Bh^J\partial_J f)=\widetilde\sG_{IJ}\Bh^J\,,\qquad\widetilde\sG_{IJ}=(\sG-\stL_{\Bh,\tilde\rho}\lsp h)_{IJ}\,,
    \label{eq:twod_symmetricflow}
\end{equation}
where $\widetilde\sG_{IJ}$ is manifestly symmetric. We now show that the sphere Ward identity guarantees that $\sw_I$ is removable.

To evaluate the sphere Ward identity, we set $\sigma=Y$ and $g^I(x)=g^I+z^IY(x)$, as in (\ref{eq:AnomOmegaDef}). On a sphere of radius $r$ we have $R=2/r^2$, $V_2=4\pi r^2$ and $\int\dd v\,\partial^\mu Y\partial_\mu Y=2N_Y/r^2$. The part of the anomaly linear in $z^I$ is then
\begin{equation}
    \mathcal{A}_Y=\frac{4N_Y}{V_2}\lsp z^I\big(\partial_I\sC-2\lsp\sw_I\big)+\text{O}(z^2)\,.
    \label{eq:twod_anomaly_side}
\end{equation}
This is linear in $\sw_I$. Substituting it into (\ref{eq:SphereWardLinear}) with $d=2$ shows that $\sw_I$ is removable,
\begin{equation}
    \sw_I=\partial_If+h_{IJ}\Bh^J\,,\qquad f=\frac12\lsp\sC-\frac{1}{8}(2+\Bh^I\partial_I)\ln Z\,,\qquad h_{IJ}=\frac{V_2}{8}(H^{(0)}-H^{(1)})_{IJ}\,.
    \label{eq:twod_w}
\end{equation}
These are regular functions of the couplings, so we can remove $\sw_I$ with local counterterms.\footnote{To define the counterterms, we work at fixed $\mu r$. The choice is arbitrary, and any dependence of $f$ and $h_{IJ}$ on $\mu r$ drops out of $\sw_I$.} With $f$ and $h_{IJ}$ of (\ref{eq:twod_w}), the condition (\ref{eq:twod_symmetricflow}) is therefore a gradient condition, with a regular potential and a regular symmetric metric.

This gradient condition coincides with that of section \ref{sec:anom}. Two relations follow from the local renormalisation group equation with constant $\sigma$. The first is (\ref{eq:radiusRG}). For the second, we differentiate the same equation twice with respect to the couplings, first along constant shifts and then along $z^IY$, and subtract the two results. The terms with $\tilde\rho_I$ then involve $\langle J^\mu\mathcal{O}_I\rangle$, which the flavour Ward identity (\ref{eq:flavourWard}) turns into the terms with $\mathcal R_I{}^J$ in $\stL_{\Bh,\tilde\rho}$. The two relations are
\begin{equation}
    (r\partial_r+\Bh^I\partial_I)\ln Z=4\sC\,,\qquad(r\partial_r+\stL_{\Bh,\tilde\rho})\lsp h_{IJ}=\sG_{IJ}\,.
\end{equation}
They imply that
\begin{equation}
    f=\frac{1}{8}(r\partial_r-2)\ln Z\,,\qquad \widetilde\sG_{IJ}=r\partial_r h_{IJ}\,.
\end{equation}
Substituting these into (\ref{eq:twod_symmetricflow}), we find that its potential and metric are
\begin{equation}
    \sC+\Bh^I\partial_If=\frac{1}{4}\left(1+\frac{1}{2}\Bh^I\partial_I\lnsp\right)r\partial_r\ln Z=\frac{A}{2\pi}\,,\qquad\widetilde\sG_{IJ}=\frac{1}{8}\lsp r\partial_r[V_2(H^{(0)}-H^{(1)})]_{IJ}=\frac{G_{IJ}}{2\pi}\,,
\end{equation}
where $A$ and $G_{IJ}$ are the limits $d\to2$ of (\ref{eq:AgradDef}). The factor of $2\pi$ reflects the normalisation $\sC=c/12$, compared with $A_*=\pi c/6$. Thus, in two dimensions, the construction of section \ref{sec:anom} from the scale response of the sphere Ward identity is equivalent to supplementing the Weyl consistency conditions with the same Ward identity. The perturbative positivity of the metric, and the relation between $A$ at fixed points and the central charge $c$, then follow as in section \ref{sec:anom}.

If the identity is the only operator of dimension zero in the reference CFT, no relevant scalar source can have a Weyl variation containing curvature or coupling derivatives. We can thus consistently set such sources to zero, without the arguments of section \ref{sec:relsource}. The gradient condition (\ref{eq:twod_symmetricflow}) is therefore a perturbative realisation of gradient flow in two dimensions.

\paragraph{Conformal manifolds.} The removability of $\sw_I$ was observed previously for theories with conformal manifolds \cite{Gomis:2015yaa}.\footnote{To translate between the conventions used here and the conventions in that paper, we note that the couplings are related by $\lambda^I=\pi g^I$, and that the anomaly coefficients are related by $\sC=c/12$, $\sG_{IJ}=\tfrac{\pi^2}{2}\lsp g_{IJ}$ and $\sw_I=-\pi^2 V_I$, where $g_{IJ}$ is the two-point function coefficient of marginal operators $\mathcal{O}_I$ spanning the conformal manifold.} There, an examination of the three-point function $\langle T_{\mu\nu} \mathcal{O}_I \mathcal{O}_J\rangle$ with exactly marginal $\mathcal{O}_I$ showed that the anomaly $\sw_I$ on the conformal manifold can only receive contributions with zero curl,
\begin{equation}
    \partial_I \sw_J-\partial_J \sw_I=0\,.
\end{equation}
The same conclusion follows from the sphere Ward identity. Suppose that $\Bh^I=0$ on an open region $\mathcal{M}$ of coupling space. On $\mathcal{M}$, (\ref{eq:twod_consistency}) implies that $\partial_I\sC=-\mathcal R_I{}^J\sw_J$, while (\ref{eq:twod_w}) gives $\sw_I=\partial_If$, so that
\begin{equation}
    \partial_I\sC=-(\tilde{\rho}_I g)^J\partial_J f=0\,.
\end{equation}
Here the derivatives $\partial_I$ are taken only along the marginal directions. The last equality holds because $f$ is flavour invariant, so that $(\omega g)^J\partial_Jf=0$ for every $\omega$ in the Lie algebra of $G$, including $\omega=\tilde\rho_I$. The central charge is thus constant on the conformal manifold, and (\ref{eq:twod_w}) becomes
\begin{equation}
    \sw_I=-\tfrac14\lsp\partial_I\ln Z\,,
    \label{eq:twod_conformal}
\end{equation}
so that $(\dd \sw)_{IJ}=0$ automatically. The relation (\ref{eq:twod_w}) extends this equation away from conformal manifolds, with both sides shifting consistently under (\ref{eq:twod_counterterm}). That this anomaly is the derivative of $\ln Z$ agrees with the explicit analysis of $\mathcal{N}=(2,2)$ theories in \cite{Gomis:2015yaa}.

\subsection{Four dimensions (and nearby)}\label{sec:fourd}
We now turn to $d=4-\ve$. The four-dimensional case follows from the smooth limit $\ve\to0$. To organise the perturbative series, we assign degree one to each coupling $g^I$ and degree two to $\ve$. The degree of a quantity is the lowest degree among its terms. In particular, $\Bh^I$ has degree two. Following Jack and Osborn~\cite{Jack:2013sha}, we work in a scheme with no $\nabla^2\sigma$ terms in the anomaly. The terms of the anomaly that we need are then those of (2.60) of \cite{Jack:2013sha}, with the four-dimensional densities continued to $d$ dimensions,
\begin{equation}\label{eq:d4-anomaly}
    \begin{split}
        \mathcal{A}_\sigma={}&-\frac{\mu^{-\ve}}{16\pi^2}\int\dd v\,\sigma\Big[\frac14\sA\cE_4+\frac1{72}\sB R^2+E^{\mu\nu}\sG_{IJ}D_\mu g^I D_\nu g^J\\
        &\hspace{30mm}+\frac{R}{6}\big(\sE_I D^2g^I+\sF_{IJ}D_\mu g^I D^\mu g^J\big)-\frac12\sA_{IJ}D^2g^I D^2g^J+\cdots\Big]\\
        &-\frac{\mu^{-\ve}}{8\pi^2}\int\dd v\,\partial_\mu\sigma\Big[E^{\mu\nu}\sw_I D_\nu g^I+\frac{R}{6}\sH_I D^\mu g^I+\sS_{IJ}D^\mu g^I D^2g^J\\
        &\hspace{30mm}+\sT_{IJK}D^\mu g^I D_\nu g^J D^\nu g^K+\mathcal{F}^{\mu\nu}\cdot\sQ_I D_\nu g^I\Big]\,,
    \end{split}
\end{equation}
where $E^{\mu\nu}=R^{\mu\nu}-\frac12\gamma^{\mu\nu}R$ is the Einstein tensor, $\cE_4=R_{\mu\nu\rho\sigma}R^{\mu\nu\rho\sigma}-4R_{\mu\nu}R^{\mu\nu}+R^2$, and $\mathcal{F}_{\mu\nu}$ is the field strength of $a_\mu$. At $d=4$, $\cE_4$ is the Euler density, but its integral is not topological for $\ve\neq0$. The terms we have omitted contain the Weyl tensor, field strengths, or at least three differentiated couplings, and do not contribute to the sphere evaluations below. As in section \ref{sec:anom}, the factor $\mu^{-\ve}$ makes the integrals dimensionless. The tensors $\sG_{IJ}$, $\sF_{IJ}$ and $\sA_{IJ}$ are symmetric, while $\sT_{IJK}=\sT_{I(JK)}$, and $\sS_{IJ}$ need not be symmetric. The tensor $\sA_{IJ}$ is an independent anomaly coefficient, unrelated to the derivatives of $\sA$. For a CFT at $d=4$, $\sA$ is the $a$-anomaly coefficient, normalised so that a free real scalar has $\sA=1/90$. The integrated anomaly of section \ref{sec:anom} is then $\mathcal{A}_*=-\sA$.

The Weyl consistency conditions that we need are derived in appendix~\ref{app:consistency}. They read
\begin{gather}
    (d-3)\lsp\partial_I\sA=\sG_{IJ}\Bh^J-(\stL_{\Bh,\tilde\rho}\lsp\sw)_I+\ve\lsp\sw_I\,,\label{eq:d4-WZ-a}\\
    \tfrac{d-1}{3}\lsp\sE_I=-\sA_{IJ}\Bh^J\,,\qquad\tfrac{d-1}{3}\lsp\sH_I=\sS_{IJ}\Bh^J\,,\label{eq:d4-WZ-EH}\\
    \tfrac{d-2}{2}\lsp(\partial_I\sw_J-\partial_J\sw_I)=-(\widehat{\sS}_{IJ}-\widehat{\sS}_{JI})\,,\label{eq:d4-WZ-curl}
\end{gather}
where
\begin{equation}\label{eq:d4-Shat}
    \widehat{\sS}_{IJ}=\tfrac{d-2}{2}\lsp\sS_{IJ}+\sS_{IK}\partial_J\Bh^K+\sT_{IJK}\Bh^K+\tfrac12\sS_{IK}\mathcal{R}_J{}^K+\tfrac12\tilde\rho_J\cdot\sQ_I\,.
\end{equation}
As in two dimensions, the first condition would be a gradient condition without the terms with $\sw_I$. Using again $(\stL_{\Bh}\sw)_I=\partial_I(\sw_J\Bh^J)-(\partial_I\sw_J-\partial_J\sw_I)\Bh^J$, we can rewrite (\ref{eq:d4-WZ-a}) as
\begin{equation}
    \partial_I\widetilde\sA=(\sG_{IJ}+\partial_I\sw_J-\partial_J\sw_I)\Bh^J+\ve \sw_I-\mathcal{R}_I{}^J\sw_J\,,\qquad \widetilde{\sA}=(d-3)\sA+\sw_I\Bh^I\,.
\end{equation}
At $d=4$, contracting with $\Bh^I$ and using $\tilde\rho_I\Bh^I=0$ gives the strong monotonicity condition
\begin{equation}
    \mu\frac{\dd}{\dd\mu}(\sA+\sw_I\Bh^I)=\Bh^I\sG_{IJ}\Bh^J\geq0\,.
\end{equation}
Away from four dimensions it becomes
\begin{equation}
    \mu\frac{\dd}{\dd\mu}\widetilde{\sA}=\Bh^I\sG_{IJ}\Bh^J+\varepsilon \sw_I\Bh^I\,,
\end{equation}
so that $\sw_I$ appears to spoil even monotonicity.

As in two dimensions, we can shift $\sw_I$ by adding the finite counterterms\footnote{These counterterms may take us out of the scheme with no $\nabla^2\sigma$ terms. Further finite counterterms return us to such a scheme without altering the shifts of $\sA$, $\sw_I$ and $\sG_{IJ}$.}
\begin{equation}\label{eq:d4-counterterm}
    \ln Z\to\ln Z-\frac{\mu^{-\ve}}{16\pi^2}\int\dd v\,\Big(\frac14\lsp b\lsp\cE_4-c_{IJ}E^{\mu\nu}D_\mu g^I D_\nu g^J\Big)\,,\qquad c_{IJ}=c_{JI}\,,
\end{equation}
where $b$ and $c_{IJ}$ are regular functions of the couplings. The anomaly coefficients then shift by
\begin{equation}\label{eq:d4-shifts}
    \sA\to\sA+(\Bh^I\partial_I-\ve)\lsp b\,,\qquad\sw_I\to\sw_I-(d-3)\lsp\partial_Ib-c_{IJ}\Bh^J\,,\qquad\sG_{IJ}\to\sG_{IJ}-(\stL_{\Bh,\tilde\rho}\lsp c)_{IJ}+\ve\lsp c_{IJ}\,.
\end{equation}
Thus, if $\sw_I$ is removable, so that $\sw_I=\partial_If+h_{IJ}\Bh^J$, choosing $(d-3)\lsp b=f$ and $c_{IJ}=h_{IJ}$ removes $\sw_I$ and turns (\ref{eq:d4-WZ-a}) into a gradient condition.

The sphere Ward identity again shows that $\sw_I$ is removable. We evaluate the anomaly (\ref{eq:d4-anomaly}) on a sphere of radius $r$ with $a_\mu=0$, $\sigma=Y$ and $g^I(x)=g^I+z^IY(x)$. Using $\nabla^2Y=-dY/r^2$ and
\begin{equation}\label{eq:d4-sphere-curvatures}
    R=\frac{d(d-1)}{r^2}\,,\qquad E^{\mu\nu}=-\frac{(d-1)(d-2)}{2r^2}\lsp\gamma^{\mu\nu}\,,\qquad\cE_4=\frac{d(d-1)(d-2)(d-3)}{r^4}\,,
\end{equation}
we find that the one-form $\Omega_I$ of (\ref{eq:AnomOmegaDef}) is
\begin{equation}\label{eq:d4-linear-anomaly}
    \begin{split}
        \Omega_I=\frac{d}{\kappa V_d}\Big[&(d-1)(d-2)\lsp\sw_I+\tfrac16d(d-1)(\sE_I-2\sH_I)\\
        &-\partial_I\big(\tfrac14(d-1)(d-2)(d-3)\lsp\sA+\tfrac1{72}d(d-1)^2\sB\big)\Big]\,,
    \end{split}
\end{equation}
where $\kappa=16\pi^2\mu^\ve r^4/V_d$, which equals $6$ at $d=4$. Using (\ref{eq:d4-WZ-EH}), the sphere Ward identity (\ref{eq:SphereWardLinear}) then gives
\begin{equation}\label{eq:d4-sphere}
    (d-1)(d-2)\lsp\sw_I-d\lsp\sS_{IJ}\Bh^J=\partial_I K+X_{IJ}\Bh^J\,,
\end{equation}
where
\begin{equation}
    \begin{aligned}
        K&=\frac{\kappa}{d}(d+\Bh^I\partial_I)\ln Z+\tfrac14(d-1)(d-2)(d-3)\lsp\sA+\tfrac1{72}d(d-1)^2\sB\,,\\
        X_{IJ}&=\tfrac d2\lsp\sA_{IJ}-\frac{\kappa}{d}V_d(H^{(0)}-H^{(1)})_{IJ}\,.
    \end{aligned}
\end{equation}
The scalar $K$ and the symmetric tensor $X_{IJ}$ are regular by the assumptions on $\ln Z$. Unlike in two dimensions, the term $\sS_{IJ}\Bh^J$ in (\ref{eq:d4-sphere}) is not manifestly removable. To show that it is, we use the condition (\ref{eq:d4-WZ-curl}).

Let $\cN$ be the set of removable one-forms (\ref{eq:removable}). We write $e\equiv e'$ when $e_I-e'_I$ belongs to $\cN$. We also define the operator $\cJ$ and the one-form $\ss_I$ by
\begin{equation}\label{eq:d4-J}
    (\cJ e)_I=(\partial_I\Bh^J)\lsp e_J\,,\qquad \ss_I=\sS_{IJ}\Bh^J\,.
\end{equation}
The operator $\cJ$ preserves $\cN$, since
\begin{equation}\label{eq:d4-J-lift}
    \cJ(\partial f+h\Bh)=\partial f'+h'\Bh\,,
\end{equation}
with
\begin{equation}\label{eq:Jshift}
    f'=\Bh^I\partial_I f+\tfrac12h_{KL}\Bh^K\Bh^L\,,\qquad h'_{IJ}=-\partial_I\partial_Jf-\tfrac12(\partial_Ih_{JK}+\partial_Jh_{IK}-\partial_Kh_{IJ})\Bh^K\,.
\end{equation}
We also need the identities
\begin{equation}\label{eq:d4-mod-identities}
    \begin{aligned}
        (\partial_I\sw_J-\partial_J\sw_I)\Bh^J&\equiv-2(\cJ \sw)_I\,,\\
        \sS_{IK}\Bh^J\partial_J\Bh^K&\equiv(\cJ \ss)_I\,,\\
        t_{IJK}\Bh^J\Bh^K&\equiv0\qquad(t_{IJK}=t_{I(JK)})\,.
    \end{aligned}
\end{equation}
For the first identity, we add $2\cJ\sw$ and write the result as $2\partial_I(\sw_J\Bh^J)-(\partial_I\sw_J+\partial_J\sw_I)\Bh^J$. For the second, we subtract $\cJ \ss$. The result is $(\sS_{IK}\partial_J\Bh^K+\sS_{JK}\partial_I\Bh^K)\Bh^J-\cJ[(\sS+\sS^T)\Bh]_I$, which is removable by (\ref{eq:d4-J-lift}). The last identity follows by writing the contraction with the symmetric tensor $(t_{IJK}+t_{JIK}-t_{KIJ})\Bh^K$.

Contracting the Weyl consistency condition (\ref{eq:d4-WZ-curl}) with $\Bh^J$ gives
\begin{equation}
    \frac{d-2}{2}\lsp(\partial_I\sw_J-\partial_J\sw_I)\Bh^J=-(d-2)\ss_I-2\sS_{IK}\Bh^J\partial_J\Bh^K-2\sT_{IJK}\Bh^J\Bh^K+2\widehat{\sS}_{(IJ)}\Bh^J\,.
\end{equation}
The last term is removable, since $\widehat{\sS}_{(IJ)}$ is symmetric. Using the identities (\ref{eq:d4-mod-identities}), we find
\begin{equation}
    (d-2)\lsp\cJ\sw\equiv(d-2)\lsp\ss+2\lsp\cJ\ss\,.
\end{equation}
Meanwhile, the sphere Ward identity (\ref{eq:d4-sphere}) gives $d\lsp\ss\equiv(d-1)(d-2)\lsp\sw$. Eliminating $\ss$ from these two relations gives
\begin{equation}
    (d-1+\cJ)\lsp\sw\equiv0\,.
\end{equation}
Hence $(d-1+\cJ)\lsp\sw=\partial f+h\Bh$ for some regular $f$ and symmetric $h_{IJ}$. Since $\Bh^I$ has degree two, $\cJ$ raises the degree of one-forms by at least one. Since also $d-1=3-\varepsilon\neq0$, the operator $d-1+\cJ$ has the inverse
\begin{equation}\label{eq:d4-invert}
    (d-1+\cJ)^{-1}=\frac{1}{d-1}\sum_{n\geq0}\left(-\frac{\cJ}{d-1}\right)^n\,,
\end{equation}
in which only finitely many terms contribute at each order. We thus find
\begin{equation}\label{eq:removablew}
    \sw=(d-1+\cJ)^{-1}(\partial f+h\Bh)\,.
\end{equation}
If $\cJ^n(\partial f+h\Bh)=\partial f^{(n)}+h^{(n)}\Bh$, as given by (\ref{eq:Jshift}), then the counterterms that remove $\sw$ are
\begin{equation}\label{eq:d4-counterterms}
    b=\frac{1}{(d-1)(d-3)}\sum_{n\geq0}\frac{(-1)^n}{(d-1)^n}f^{(n)}\,,\qquad c_{IJ}=\frac{1}{d-1}\sum_{n\geq 0}\frac{(-1)^n}{(d-1)^n}h^{(n)}_{IJ}\,.
\end{equation}
These counterterms are regular. The construction respects flavour covariance, so $b$ can be taken invariant. Consequently, $\sw_I$ is removable.

The counterterms (\ref{eq:d4-counterterms}) are complicated, because of the inverse in (\ref{eq:removablew}). Nevertheless, they preserve the matching of the potential with $\widetilde{F}$ and the perturbative positivity of the metric found in section \ref{sec:anom}. Keeping track of the removable terms, one finds that $f_*=K_*/(d-2)$ at a fixed point. Since $\sB$ vanishes at fixed points,
\begin{equation}
    K_*=\kappa\ln Z|_{g_*}+\frac{(d-1)(d-2)(d-3)}{4}\sA_*\,.
\end{equation}
By (\ref{eq:Jshift}), $f^{(n)}$ vanishes at fixed points for $n\geq 1$, so only the term with $n=0$ contributes to $b$ there, and
\begin{equation}
   b_*=\frac{1}{(d-1)(d-2)(d-3)} K_*\,.
\end{equation}
In the new scheme $\sw_I=0$, and (\ref{eq:d4-WZ-a}) is a gradient condition with potential $(d-3)\sA$. By (\ref{eq:d4-shifts}), its value at a fixed point is
\begin{equation}
    (d-3)(\sA_*-\ve\lsp b_*)=-\frac{\kappa}{(d-1)(d-2)}(\mathcal{A}_*+\varepsilon\ln Z|_{g_*})\,,
\end{equation}
where $\mathcal{A}_*=-d(d-1)(d-2)(d-3)\lsp\sA_*/(4\kappa)$ is the integrated anomaly at the fixed point in $d$ dimensions. By (\ref{eq:ZtoZCFT}), $\mathcal{A}_*+\varepsilon\ln Z|_{g_*}=\varepsilon\ln Z_{\text{CFT}}$, where $\ln Z_{\text{CFT}}$ is the radius-independent sphere free energy that defines $\widetilde{F}$. Hence
\begin{equation}
    (d-3)(\sA_*-\ve\lsp b_*)=-\frac{\kappa \varepsilon}{(d-1)(d-2)\sin\frac{\pi d}{2}}\widetilde{F}_*\,.
\end{equation}
The prefactor is positive near $d=4$, in agreement with section \ref{sec:anom}. The shift of $\sG_{IJ}$ in (\ref{eq:d4-shifts}) has positive degree for regular $c_{IJ}$, so it does not affect the leading term of the metric. The positivity of the leading term of $\sG_{IJ}$ therefore still guarantees the positivity of the shifted metric.

Relevant scalar sources were omitted above. Near four dimensions these are the mass sources $M$ of dimension two, whose index we suppress. Section \ref{sec:relsource} eliminates them from the gradient condition of section \ref{sec:anom}. The removal of $\sw_I$ by counterterms is a statement about the anomaly, however, and the anomaly depends on the form of the Weyl variation of the sources. We therefore treat the mass sources explicitly here.

Redefinitions of $M$ by multiples of $R$ and of $D^2g^I$ shift the coefficients of $\nabla^2\sigma$ and of $\partial^\mu\sigma\lsp D_\mu g^I$ in $\Delta_\sigma M$ at leading order by multiples of $2(d-1)$ and $d-2$. Order by order, they lead to a scheme in which $\Delta_\sigma M$ contains no derivatives of $\sigma$,
\begin{equation}\label{eq:mass-variation}
    \Delta_\sigma M=-\sigma\big[(2-\gamma_M)\lsp M+m\big]\,,\qquad m=\eta\lsp R+\delta_ID^2g^I+\epsilon_{IJ}D_\mu g^ID^\mu g^J\,,
\end{equation}
as at $d=4$ \cite{Jack:2013sha}. The matrix $\gamma_M$ contains the anomalous dimensions of the mass sources, and $\eta$, $\delta_I$ and $\epsilon_{IJ}=\epsilon_{JI}$ are functions of the couplings. Commutativity of Weyl rescalings on $M$ requires $\sigma'\Delta_\sigma m=\sigma\Delta_{\sigma'}m$, so that $\Delta_\sigma m$ contains no derivatives of $\sigma$ either. The absence of terms proportional to $\nabla^2\sigma$ and to $\partial^\mu\sigma\lsp D_\mu g^I$ gives
\begin{equation}\label{eq:mass-relations}
    \delta_I\Bh^I=2(d-1)\lsp\eta\,,\qquad\big[\big(\tfrac{d-2}{2}+\cJ+\tfrac12\mathcal R\big)\lsp\delta\big]_I=-\epsilon_{IJ}\Bh^J\,,
\end{equation}
where $\cJ$ is the operator of (\ref{eq:d4-J}) and $\mathcal R$ acts on one-forms as $(\mathcal Re)_I=\mathcal R_I{}^Je_J$. At $d=4$ these relations reduce to the consistency conditions of \cite{Jack:2013sha}. Both $\cJ$ and $\mathcal R$ raise the degree. The second relation therefore shows that $\delta_I$ has degree at least two, and the first that $\eta$ has degree at least four.

The anomaly (\ref{eq:d4-anomaly}) then also contains terms linear and quadratic in $M$. We use finite counterterms of the form $\int\dd v\,M\lsp(j\lsp R+u_ID^2g^I)$ to choose a scheme in which these terms contain no derivatives of $\sigma$ or of $M$. The two possible terms with such derivatives, $M\nabla^2\sigma$ and $M\lsp\partial^\mu\sigma D_\mu g^I$, are shifted at leading order by $-2(d-1)\lsp j$ and $(d-2)\lsp u_I$. In this scheme the terms of the anomaly that contain $M$ are
\begin{equation}\label{eq:mass-anomaly}
    \frac{\mu^{-\ve}}{16\pi^2}\int\dd v\,\sigma\Big[-\frac R6\lsp\sI\lsp M+\frac12\lsp\sP M^2+\sJ_IM\lsp D^2g^I+\sK_{IJ}M\lsp D_\mu g^ID^\mu g^J\Big]\,,
\end{equation}
where $\sI$, $\sP$, $\sJ_I$ and $\sK_{IJ}=\sK_{JI}$ are functions of the couplings. The coefficient $\sJ_I$ is unrelated to the operator $\cJ$ and to the currents $J^\mu$. Commutativity of Weyl rescalings on the terms of the anomaly that are linear in $M$ gives, in the same way as (\ref{eq:mass-relations}),
\begin{equation}\label{eq:mass-anomaly-relations}
    \sJ_I\Bh^I=-\tfrac{d-1}{3}\lsp\sI\,,\qquad\big[\big(\tfrac{d-2}{2}+\cJ+\tfrac12\mathcal R\big)\lsp\sJ\big]_I=-\sK_{IJ}\Bh^J\,.
\end{equation}

On the sphere we choose the source configuration
\begin{equation}\label{eq:mass-configuration}
    M=R\lsp\nu(g)\,,\qquad(2-\gamma_M+\Bh^I\partial_I)\lsp\nu=-\eta\,.
\end{equation}
The operator acting on $\nu$ has the invertible leading term $2$, so that $\nu$ is unique and regular. Like $\eta$, it has degree at least four and vanishes at fixed points. In the sphere Ward identity, the couplings change by $Y\Bh^I-\xi^\mu\partial_\mu g^I$ and the sources by $\Delta_YM-\xi^\mu\partial_\mu M$. On the configuration (\ref{eq:mass-configuration}), this change of the sources equals the change of $R\lsp\nu(g)$ induced by the change of the couplings, up to the term $-Y\varrho$, where $\varrho=\delta_ID^2g^I+\epsilon_{IJ}D_\mu g^ID^\mu g^J$ contains the derivatives of the couplings. Substituting (\ref{eq:mass-configuration}) into $\ln Z$ therefore gives a functional of the metric and the couplings alone, which obeys (\ref{eq:sphereWardAnom}) with the additional term $-\int\dd^d x\,Y\varrho\,\delta\ln Z/\delta M$ on the left-hand side.

With these choices the sources enter the removal of $\sw_I$ in three places. First, the consistency conditions (\ref{eq:d4-WZ-a})--(\ref{eq:d4-WZ-curl}) are unchanged. A term of the anomaly that depends on $M$ can contribute to a condition without sources only through the part $-\sigma\lsp m$ of $\Delta_\sigma M$. Neither this part nor the term itself contains derivatives of $\sigma$. The contribution is therefore symmetric under the exchange of $\sigma$ and $\sigma'$, and cancels in $\Delta_\sigma\mathcal{A}_{\sigma'}-\Delta_{\sigma'}\mathcal{A}_\sigma$. Second, the Ward identity gains the additional term. At first order in $z$ we have $\varrho=-(d/r^2)\lsp z^I\delta_IY$, while $\delta\ln Z/\delta M$ is evaluated at constant couplings, where it equals $\sqrt\gamma\lsp\chi$ with $\chi$ constant on the sphere. The right-hand side of (\ref{eq:SphereWardLinear}) thus becomes $\Omega_I-\sf_I\chi$, with $\sf_I=d\lsp\delta_I/r^2$. Third, on the configuration (\ref{eq:mass-configuration}) the terms (\ref{eq:mass-anomaly}) add to the one-form $\Omega_I$ the contribution
\begin{equation}\label{eq:mass-Omega}
    \Omega^M_I=\frac{\mu^{-\ve}R}{16\pi^2}\Big[\partial_I\Big(\frac R2\lsp\sP\nu^2-\frac R6\lsp\sI\lsp\nu\Big)-\frac{d}{r^2}\lsp\nu\lsp\sJ_I\Big]\,.
\end{equation}
The term with $\sK_{IJ}$ is of second order in $z$ and does not contribute. The right-hand side of (\ref{eq:d4-sphere}) thus gains
\begin{equation}\label{eq:d4-sphere-sources}
    \frac{\kappa V_d}{d}\lsp\big(\sf_I\chi-\Omega^M_I\big)\,.
\end{equation}

These terms are removable. The operator in (\ref{eq:mass-relations}) and (\ref{eq:mass-anomaly-relations}) acts algebraically on one-forms, so it commutes with multiplication by $\chi$ and by $\nu$, and
\begin{equation}\label{eq:mass-removal}
    \begin{aligned}
        \big[\big(\tfrac{d-2}{2}+\cJ+\tfrac12\mathcal R\big)(\chi\lsp\delta)\big]_I&=-\chi\lsp\epsilon_{IJ}\Bh^J\,,\\
        \big[\big(\tfrac{d-2}{2}+\cJ+\tfrac12\mathcal R\big)(\nu\lsp\sJ)\big]_I&=-\nu\lsp\sK_{IJ}\Bh^J\,.
    \end{aligned}
\end{equation}
The right-hand sides are removable, since $\chi\lsp\epsilon_{IJ}$ and $\nu\lsp\sK_{IJ}$ are regular and symmetric. The operator $\cJ$ preserves $\cN$ by (\ref{eq:d4-J-lift}). So does $\mathcal R$, because $\mathcal R_I{}^J\partial_Jf=0$ for flavour-invariant $f$ and $\Bh^I\mathcal R_I{}^J=0$, so that $\mathcal R$ maps $h_{IJ}\Bh^J$ to $(\mathcal R_I{}^Kh_{KJ}+\mathcal R_J{}^Kh_{KI})\Bh^J$. As in (\ref{eq:d4-invert}), the operator on the left-hand sides of (\ref{eq:mass-removal}) therefore has a regular inverse that preserves $\cN$, so that $\sf_I\chi$ and $\nu\lsp\sJ_I$ are removable. The rest of $\Omega^M_I$ is the gradient of a regular and flavour-invariant function, which is removable as well. Hence (\ref{eq:d4-sphere}) continues to hold up to removable terms, and $\sw_I$ remains removable.

The sources affect neither the value of the potential at fixed points nor the leading metric. They vanish at fixed points, since $\nu$ does. The scalar parts of the additional removable terms vanish there as well, because they contain a factor of $\nu$ or two factors of $\Bh^I$. The additional contributions to $c_{IJ}$ are regular, so by (\ref{eq:d4-shifts}) they shift $\sG_{IJ}$ only at positive degree.

\paragraph{Gauge couplings.} In four dimensions, gauge couplings are also marginal, and the arguments of section \ref{sec:anom} and of this section extend to them with one modification. For a gauge group factor $\mathcal{G}_a$ with field strength $F_a$ and kinetic term $F_a^2/(4\gs_a^2)$, we use the coordinate $\lambda^a=\log\gs_a$, with $\partial_{\lambda^a}=\gs_a\partial_{\gs_a}$ and
\begin{equation}\label{eq:gauge-log}
    \Bh^{\lambda^a}=\frac{\Bh^{\gs_a}}{\gs_a}=-\frac{\ve}{2}+\text{O}(g^2)\,,
\end{equation}
where the terms of order $g^2$ are loop corrections. We keep the theta angles fixed and exclude kinetic mixing between abelian factors. Flavour rotations do not act on gauge couplings. Together with the other couplings, the $\lambda^a$ form the coordinates $g^I$. We still write $g^I=0$ for the free theory, although it lies at $\lambda^a=-\infty$.

In these coordinates $\Bh^{\lambda^a}$ still has degree two, since the constant $-\ve/2$ in (\ref{eq:gauge-log}) has degree two. The derivative $\partial_{\lambda^a}$ does not lower the degree, so $\cJ$, $\Bh^I\partial_I$ and $\stL_{\Bh,\tilde\rho}$ still raise it. The degree arguments above therefore apply to gauge couplings unchanged.

Free Yang--Mills theory tests the leading metric in these coordinates. On the sphere, its constant gauge transformations are ghost zero modes. For a simple gauge group, they give $\ln Z=\dim\mathcal{G}\,\lambda+\ln Z_0(\mu r)$, where $Z_0$ does not depend on $\gs$ \cite{DeCesare:2023sphere}. With $\Bh^{\lambda}=-\ve/2$ exactly, (\ref{eq:AgradDef}) gives $\partial_\lambda A=\sin\frac{\pi d}{2}\,\dim\mathcal{G}$. The gradient condition then requires
\begin{equation}\label{eq:gauge-metric}
    G_{\lambda\lambda}=-\frac{2}{\ve}\sin\frac{\pi d}{2}\,\dim\mathcal{G}=\pi\dim\mathcal{G}+\text{O}(\ve^2)\,.
\end{equation}
The two-point function reproduces this value at $d=4$. The operator conjugate to $\lambda$ is $\mathcal{O}_\lambda=-\frac12F_{\text{c}}^2$, with $F_{\text{c}}=F/\gs$, and $\langle F_{\text{c}}^2(x)\lsp F_{\text{c}}^2(0)\rangle=48/(\pi^4x^8)$ for each generator. Hence $C_{\lambda\lambda}=12\dim\mathcal{G}/\pi^4$, and the leading metric of section \ref{sec:anom} is $G_{\lambda\lambda}|_{g=0}=\frac{2\pi^5}{4!}\lsp C_{\lambda\lambda}=\pi\dim\mathcal{G}$, which is constant and positive in $\lambda$. In the coordinate $\gs$ it equals $\pi\dim\mathcal{G}/\gs^2$, in agreement with the dependence on $1/\gs^2$ found by Jack and Osborn \cite{Jack:1990eb}.

The one modification is that the potential and the counterterms that remove $\sw_I$ contain logarithms of the gauge couplings. In free Yang--Mills theory, the potential contains the term $\sin\frac{\pi d}{2}\,\dim\mathcal{G}\log\gs=-\frac{\ve}{2}\lsp G_{\lambda\lambda}\log\gs$, as a constant metric requires, and this term vanishes at $d=4$. At $d=4$ the free theory has $\Bh^\lambda=0$, and (\ref{eq:d4-sphere}) with $\kappa=6$ gives $\sw_\lambda=\dim\mathcal{G}$. This one-form is removable with $f=\dim\mathcal{G}\log\gs$, so the counterterm $b$ of (\ref{eq:d4-counterterm}) contains a logarithm even at $d=4$. In the coordinates $\lambda^a$ these logarithms are linear functions. With gauge couplings, the regular quantities of section \ref{sec:intro} are therefore perturbative series in $\gs_a$ and in the other couplings, with coefficients that are polynomials in $\lambda^a$. The operators $\cJ$, $\Bh^I\partial_I$ and $\stL_{\Bh,\tilde\rho}$ preserve this class of functions. The Gaussian fixed point lies at $\lambda^a=-\infty$, outside this logarithmic coordinate chart. For $\ve\neq0$, free Maxwell theory is not conformally invariant \cite{El-Showk:2011xbs}, so the matching statement for conformal fixed points does not apply to this endpoint. At $d=4$ the free theory is conformally invariant. The potential has a finite limit there, since its logarithm vanishes at $d=4$. The counterterm $b$ diverges logarithmically in this limit.

These results assume, as stated in section \ref{sec:intro}, that the anomaly is local. For gauge couplings we have not proved this assumption. The ghost zero modes could make the anomaly on the sphere non-local, because the volume of the constant gauge transformations and the ghost determinant without its zero modes each have non-local Weyl variations. Locality of the anomaly requires these two non-local variations to cancel. The check of (\ref{eq:gauge-metric}) tests the Ward identity at leading order, but not locality in general. Under this assumption, the gradient condition holds for four-dimensional theories with gauge, Yukawa and scalar couplings, whose $a$-function and consistency conditions were studied in \cite{Jack:2014pua,Poole:2019txl}.

\subsection{Six dimensions}\label{sec:sixd}
The six-dimensional local renormalisation group was developed in \cite{Grinstein:2013cka,Stergiou:2016uqq} for perturbative fixed points such as those of $\phi^3$ theory in $d=6-\ve$. In the sectors with at most two free coupling indices, the anomaly of \cite{Stergiou:2016uqq} gives nine scalar, twenty-four one-form and nineteen two-index consistency conditions. Most of them are listed in the ancillary file of that reference. In its notation, one combination of them, (3.5) of \cite{Stergiou:2016uqq}, reads $\partial_I\hat a=(\chi_{IJ}+\xi_{IJ})\beta^J$. Here $\hat a$ reduces to the Euler coefficient $a$ at fixed points, $\chi_{IJ}$ is symmetric and $\xi_{IJ}$ is antisymmetric. The antisymmetric part spoils a gradient interpretation.

The construction of section \ref{sec:anom} applies at $D=6$ without relevant sources. Its leading metric is $G_{IJ}|_{g=0}=\pi^7C_{IJ}/360$, which is positive definite in a unitary theory, so that $A$ decreases towards the infrared at weak coupling. For $\ve\to0$, its fixed-point value is $A_*=\pi\mathcal{A}_*/2$. On the round sphere only the Euler term contributes to $\mathcal{A}_*$ \cite{Deser:1993yx}, so $A_*$ is proportional to $a$.

We now compare with \cite{Stergiou:2016uqq} at $d=6$, using its conventions. Its $\beta$ is our $\Bh$, and calligraphic letters with an index $I$ denote its one-form anomaly coefficients. The comparison rests on the consistency condition $\Delta_Y\mathcal{A}_1=\Delta_1\mathcal{A}_Y$, evaluated on the sphere with $g^I(x)=g^I+z^IY(x)$ at first order in $z$. The local functional $\mathcal{A}_1=\Delta_1\ln Z$ obeys the sphere Ward identity, so the steps that led to (\ref{eq:SphereWardLinear}) apply to the left-hand side. On the right-hand side, $\Delta_1$ acts on $\mathcal{A}_Y=N_Yz^I\Omega_I$ in three ways. The scaling with the radius contributes $-\ve$, the changes $\delta g^I=\Bh^I$ and $\delta z^I=z^J\partial_J\Bh^I$ give the Lie derivative of $\Omega_I$, and the change of $a_\mu$ gives the flavour part of the extended Lie derivative. The result is
\begin{equation}\label{eq:d4-sphere-consistency}
    (\stL_{\Bh,\tilde\rho}\lsp\Omega)_I-\ve\lsp\Omega_I=\partial_I\left[\frac{1}{V_d}\left(d+\Bh^J\partial_J\right)\mathcal{A}_1\right]+(\mathcal{H}^{(1)}-\mathcal{H}^{(0)})_{IJ}\Bh^J\,,
\end{equation}
where $\mathcal{H}^{(\ell)}_{IJ}$ are defined by (\ref{eq:HIJdefn}) and (\ref{eq:angmoms}), with $\mathcal{A}_1$ in place of $\ln Z$. This relation involves only the anomaly, so we may apply it with the anomaly functional of \cite{Stergiou:2016uqq}. By (\ref{eq:SphereWardLinear}), $\Omega_I=\partial_Iu+v_{IJ}\Bh^J$, with $u=(d+\Bh^J\partial_J)\ln Z/V_d$ and $v_{IJ}=(H^{(1)}-H^{(0)})_{IJ}$. Substituting this into (\ref{eq:d4-sphere-consistency}) gives a gradient condition with potential $\Bh^I\partial_Iu-\ve u-(d+\Bh^I\partial_I)\mathcal{A}_1/V_d$ and metric $\mathcal{H}^{(1)}-\mathcal{H}^{(0)}-(\stL_{\Bh,\tilde\rho}-\ve)\lsp v$. On the unit sphere, $\mathcal{H}^{(1)}_{IJ}-\mathcal{H}^{(0)}_{IJ}=720\lsp g_{IJ}$ up to terms that depend on the flow or its derivatives. The tensor $g_{IJ}$ is that of (3.10) of \cite{Stergiou:2016uqq}, positive definite by unitarity according to that reference. The factor $720$ is the Euler density of the unit sphere in the normalisation of that reference. The potential is $6\cdot720\lsp a$ up to terms that vanish with the flow. Dividing by $6\cdot720$ gives the leading form $\partial_Ia=\tfrac16\lsp g_{IJ}\beta^J$ of (3.5) of \cite{Stergiou:2016uqq}. For the anomaly of \cite{Stergiou:2016uqq}, the one-form $\Omega_I$ of (\ref{eq:AnomOmegaDef}) is
\begin{equation}
    \begin{split}
        \Omega_I={}&-216\lsp\mathcal A_I+216\lsp\mathcal B_I+180\lsp\mathcal C_I+360\lsp\mathcal G^1_I+216\lsp\mathcal G^3_I+720\lsp\mathcal H^1_I+180\lsp\mathcal H^2_I+150\lsp\mathcal H^3_I+150\lsp\mathcal H^4_I\\
        &-\partial_I\big(720\lsp a-75\lsp b_1-45\lsp b_2-\tfrac{71}{2}\lsp b_3-90\lsp b_4+900\lsp b_5-90\lsp b_6\big)\,.
    \end{split}
    \label{eq:sixd_Omega}
\end{equation}
The one-forms in the first line are the coefficients of the anomaly terms with one differentiated coupling, and the gradient in the second line comes from the terms with no derivatives of the couplings or of $\sigma$. Its coefficients are the values of their curvature invariants on the unit sphere. The invariant multiplying $b_7$ vanishes there, as do the Weyl invariants. The one-forms $\mathcal A_I,\dots,\mathcal H^4_I$ should not be confused with the anomaly $\mathcal{A}_\sigma$ or the moments $\mathcal{H}^{(\ell)}_{IJ}$.

The relation (\ref{eq:d4-sphere-consistency}) also tests the published consistency conditions. Each one-form condition is the coefficient of one local structure in the Weyl commutator $\Delta_\sigma\mathcal{A}_{\sigma'}-\Delta_{\sigma'}\mathcal{A}_\sigma$. In the configuration used for (\ref{eq:d4-sphere-consistency}), each structure integrates to a pure number times $N_Yz^I$. The relation (\ref{eq:d4-sphere-consistency}) must therefore be a sum of the one-form conditions, each weighted by the number of its structure. No such sum of the conditions published in \cite{Stergiou:2016uqq} reproduces it, so the published conditions contain an error.

We therefore rederived the scalar and one-form conditions independently. We kept all anomaly terms with at most two differentiated couplings. We then took the functional derivative of the integrated commutator with respect to $\sigma'(x)$. The result is a local expression that must vanish identically. Unlike an integrand, it has no ambiguity from total derivatives. We evaluated it exactly at a point, for random values of the metric, the couplings, $\sigma$ and their derivatives. We did this with one and with two couplings. Each evaluation gives a linear relation among the anomaly coefficients. Terms with no differentiated coupling give the scalar conditions, and terms with one differentiated coupling give the one-form conditions. This procedure requires neither a basis of independent structures nor integration by parts. Every relation obtained in this way is an exact consequence of the commutativity of Weyl rescalings. We added evaluations until the rank of these relations stopped growing.

We recover all nine scalar conditions. We also recover twenty-two of the twenty-four one-form conditions. The two exceptions are the ninth and eleventh entries of the list in the ancillary file of \cite{Stergiou:2016uqq}, which we call $V_9$ and $V_{11}$. They are the coefficients of $(\sigma\partial_\mu\sigma'-\sigma'\partial_\mu\sigma)H_2^{\mu\nu}\partial_\nu g^I$ and $(\sigma\partial_\mu\sigma'-\sigma'\partial_\mu\sigma)H_4^{\mu\nu}\partial_\nu g^I$, respectively. The tensors $H_2$ and $H_4$ are defined in (B.4) of \cite{Stergiou:2016uqq}. We write each condition as an expression that must vanish. We normalise $V_9$ so that the Lie derivative of $\mathcal H^2_I$ has unit coefficient, and $V_{11}$ so that the Lie derivative of $\mathcal H^4_I$ has unit coefficient. The rederivation then requires the corrections
\begin{equation}
    V_9\to V_9-\tfrac52\lsp\hat{\mathcal A}''_{IJ}\beta^J-\tfrac54\lsp\partial_I(\mathcal C_J\beta^J)\,,\qquad V_{11}\to V_{11}+\hat{\mathcal A}''_{IJ}\beta^J\,,
    \label{eq:ssv_corrections}
\end{equation}
where $\hat{\mathcal A}''_{IJ}$ is the tensor of that name in \cite{Stergiou:2016uqq}. The corrections remove $\hat{\mathcal A}''_{IJ}$ from both conditions. With them, the published one-form conditions, together with the derivatives of the scalar conditions, span the same space as the rederived relations with one differentiated coupling. This holds with one and with two couplings. The sum that gives (\ref{eq:d4-sphere-consistency}) then exists.\footnote{Its weights, in the order of the ancillary list, are
\begin{equation}
    (216,216,216,-1080,-1080,0,360,-90,180,150,150,0,0,360,360,0,36,360,720,0,5400,0,0,0)\,.
    \label{eq:ssv_weights}
\end{equation}} The corrected conditions lie in the span of the rederived relations, so they hold exactly. Using the complete basis of local structures constructed in Appendix A of \cite{Stergiou:2016uqq}, we also establish their completeness. With two couplings, the rederived relations with one differentiated coupling have rank 66. This is the number of components of the twenty-four one-form conditions and of the derivatives of the nine scalar conditions, and therefore the largest rank that this basis allows.

The corrections do not affect the conclusions of \cite{Stergiou:2016uqq}. Equation (3.4) of that reference, for $\tilde a$, remains valid. Equation (3.5) of that reference, $\partial_I\hat a=(\chi_{IJ}+\xi_{IJ})\beta^J$, combines several conditions, including $V_9$ with weight $\tfrac1{20}$ and $V_{11}$ with weight $\tfrac18$. The corrections (\ref{eq:ssv_corrections}) therefore change this combination by
\begin{equation}
    \tfrac1{20}\Big(-\tfrac52\lsp\hat{\mathcal A}''_{IJ}\beta^J-\tfrac54\lsp\partial_I(\mathcal C_J\beta^J)\Big)+\tfrac18\lsp\hat{\mathcal A}''_{IJ}\beta^J=-\tfrac1{16}\lsp\partial_I(\mathcal C_J\beta^J)\,.
\end{equation}
The $\hat{\mathcal A}''$ terms cancel, and the remainder is a gradient. A gradient shifts $\hat a$ and leaves $\chi_{IJ}$ and $\xi_{IJ}$ unchanged. In the expression for $\hat a$ in (3.6) of \cite{Stergiou:2016uqq}, the term $\tfrac1{24}\lsp\mathcal C_I\beta^I$ becomes $-\tfrac1{48}\lsp\mathcal C_I\beta^I$. This term vanishes at fixed points, where $\hat a=a$, so the results for $a$ are unaffected.

Independently of these corrections, the rederivation changes two coefficients in the expression for $\chi_{IJ}$ in (3.6) of \cite{Stergiou:2016uqq}. That reference symmetrises and antisymmetrises without a factor of $\tfrac12$, so that $X_{(IJ)}=X_{IJ}+X_{JI}$ and $X_{[IJ]}=X_{IJ}-X_{JI}$, as stated in its (3.7). With this convention, the coefficients of $\hat{\mathcal E}_{(IJ)}$ and $\mathcal F_{(IJ)}$ in $\chi_{IJ}$ should be $\tfrac1{40}$ and $\tfrac1{48}$, in place of $\tfrac1{20}$ and $\tfrac1{24}$. They then equal the coefficients of $\hat{\mathcal E}_{[IJ]}$ and $\mathcal F_{[IJ]}$ in $\xi_{IJ}$, which the rederivation confirms.

For $\phi^3$ theory, section \ref{sec:relsource-invert} shows that the source of $\phi$ cannot be eliminated by the construction of section \ref{sec:relsource}. Our result for $\phi^3$ theory near six dimensions is therefore conditional on a treatment of this source, which remains open.

\section{Conclusion}\label{sec:conclusion}
In this paper we have shown that a conformal Ward identity on the sphere gives a gradient condition for perturbative RG flows in any dimension, with or without a Weyl anomaly. Central to our argument has been the existence of conformal Killing vectors on the sphere that are not isometries. The Ward identity follows by setting the action of these vectors on $\ln Z$ as diffeomorphisms equal to their action as Weyl transformations. With and without an anomaly, the resulting monotonic function $A$ matches the rescaled sphere free energy $\widetilde{F}$ of \cite{Giombi:2014xxa,Fei:2015oha} at fixed points, which establishes the weak $\widetilde{F}$-theorem to all loop orders in perturbation theory. The metric $G_{IJ}$ is manifestly symmetric. At leading order it becomes the Zamolodchikov metric of two-point function coefficients of the perturbing operators, which is positive by unitarity.

The gradient condition was first derived with marginal sources only. We have also shown that the sources of relevant operators can be eliminated by a stationarity condition. This changes the form of $A$ and $G_{IJ}$ without altering either the matching of $A$ with $\widetilde{F}$ at fixed points or the identification of the leading term in the metric with the Zamolodchikov metric. The elimination requires an invertible matrix of second derivatives with respect to the sources. In unitary theories this fails only for the source of a free scalar field near $d=6,10,\ldots\,$, where the sphere Ward identity alone does not establish gradient flow. Extensive perturbative evidence in $d=6$ and $d=6-\varepsilon$ indicates that gradient flow nevertheless remains valid~\cite{Gracey:2015fia,Benfatto:2025yci,Stergiou:phi3GraphAtlas}. This suggests that an additional argument should cover this source.

We set up the local renormalisation group for a theory with an action that is free at $g^I=0$. After section \ref{sec:localRG}, however, our derivation does not use this. It uses the local renormalisation group equation (\ref{eq:localRG}), the conformal invariance of the theory at $g^I=0$, and the assumptions of regularity and locality stated in section \ref{sec:intro}. We therefore expect our results to hold in conformal perturbation theory, where an interacting CFT is perturbed by operators of dimension $\Delta_I=d-\delta_I$ with small $\delta_I$. The local renormalisation group has been formulated in this setting \cite{Baume:2014rla,Stergiou:2016uqq}, and our argument for the long-range model in section \ref{sec:free} is of this type. At leading order the result is already known. The leading metric of section \ref{sec:free} gives $\partial_IA=2\pi^{d+1}C_{IJ}\Bh^J/\Gamma(d+1)$ at this order, which for a single operator is the relation between the sphere free energy and the beta function found in \cite{Klebanov:2011gs,Fei:2015oha}. In two dimensions it is Zamolodchikov's relation between the $c$-function and the beta function \cite{Zamolodchikov:1987ti}. Beyond leading order that relation fails if the metric is the two-point function at a fixed separation \cite{Freedman:2005wx}, and that reference asked whether another metric exists. Our metric is instead a two-point function smeared over the sphere, as in the gradient formula for boundary flows \cite{Friedan:2003yc}. It is important to point out that an interacting CFT cannot in general be continued in $d$, so in an even dimension $D$ one must work at $\varepsilon=0$. The argument of section \ref{sec:anom} then applies with $r\partial_r$ in place of $r\partial_r+\varepsilon$, as in section \ref{sec:twod}, and the potential and the metric are the limits $\varepsilon\to0$ of (\ref{eq:AgradDef}). Moreover, the assumptions of regularity and locality are harder to justify, because dimensional regularisation is not available for a general CFT. In the case of $\widetilde{F}$, it has been shown previously that it is possible to extend it away from fixed points in a gradient manner through second order in conformal perturbation theory~\cite{Fraser-Taliente:2026iuj}. Checks of our construction beyond leading order would therefore be valuable.

The compactness of the sphere was crucial for our derivation. It provides an IR regulator for the definition of objects like $\ln Z$, and it ensures that the Ward identities (\ref{eq:linWard}) and (\ref{eq:SphereWardLinear}) receive no contributions from boundary terms. On a more general manifold, boundary terms can introduce an additional one-form that is not immediately removable, which spoils any conclusion of gradient flow. Ward identities on other non-trivial manifolds may nevertheless contain valuable information. For instance, in two dimensions the $c$-theorem has been extended beyond unitary theories to $\mathcal{P}\mathcal{T}$-symmetric theories, as a $c_{\text{eff}}$-theorem~\cite{Castro-Alvaredo:2017udm}. Here $c_{\text{eff}}=c-24\Delta$, where $\Delta$ is the lowest dimension of a Virasoro primary in the CFT. Without the additional boundary one-form, the Ward identity on the cylinder would produce a monotone $A$ that agrees with $c_{\text{eff}}$ at fixed points. A deeper understanding of this boundary one-form may thus permit an extension of the $c_{\text{eff}}$-theorem to higher dimensions, or strengthen its monotonicity. Hyperbolic space, which like the sphere has conformal Killing vectors that are not isometries, is another natural place to look. Osborn and Shore analysed correlation functions of the energy-momentum tensor on it through a spectral representation \cite{Osborn:1999az}. A Ward identity there would have to account carefully for boundary terms at infinity.

Ward identities on the sphere already have a long history in connection with RG monotonicity for defects. They have been used to prove the $g$-theorem for line defects~\cite{Friedan:2003yc,Cuomo:2021rkm} and the $b$-theorem for surfaces~\cite{Shachar:2022fqk}. It would be interesting to see whether the methods used here can establish gradient flow perturbatively for defects of arbitrary dimension. The Ward identity for defects would need to include a contribution from the bulk function $\Bh^I(g)$. Indeed, a recent computation of the beta function of the scalar line defect in $d=4-\varepsilon$ through five loops~\cite{PannellDefects} indicates that it satisfies a gradient condition
\begin{equation}
    \partial_i A(h,g)=G_{ij}\Bh^{j}(h,g)+G_{iJ}\Bh^J(g)
\end{equation}
only if a non-zero mixed metric $G_{iJ}$ is included. Here $h^i$ are the couplings on the defect.

Although our work is perturbative, the questions we study also have non-perturbative forms. In $d=2$, Zamolodchikov proved strong monotonicity non-perturbatively~\cite{Zamolodchikov:1986gt}. In $d=4$, the weak $a$-theorem has been proved using dilaton scattering \cite{Komargodski:2011vj}, entanglement inequalities \cite{Casini:2017vbe} and the averaged null energy condition (ANEC) \cite{Hartman:2023qdn}. The strong $a$-theorem was also recently proved non-perturbatively using the ANEC~\cite{Hartman:2026nqa}, which also provided a non-perturbative understanding of the Jack--Osborn metric. A non-perturbative approach to the local RG has also been developed in two dimensions \cite{Friedan:2009ik}, with the antisymmetric contribution to the metric present, while in four dimensions supersymmetry allows an exact discussion of the gradient property \cite{Barnes:2004jj}. A more general understanding of the sphere Ward identity may help extend these results and establish the gradient property non-perturbatively.

\ack{We thank Ian Jack, Zohar Komargodski, Hugh Osborn, and Omar Zanusso for useful discussions. Computations and manuscript review were carried out with the help of Anthropic's Claude Opus 5.5 and Fable 5.2, and OpenAI's GPT-6 Astra. AS is supported by the Royal Society under grant URF{\textbackslash}R1{\textbackslash}211417 and by STFC under grant ST/X000753/1.}

\appendix

\section{Perturbative checks}\label{app:pertchecks}
The gradient property constrains the beta function, in that the coefficients of $\Bh^I$ must obey an infinite tower of conditions for any potential to exist. These conditions arise when we solve
\begin{equation}\label{eq:gradgeneral}
    \partial_I A(g)=G_{IJ}(g)\Bh^J(g)
\end{equation}
order by order in perturbation theory, with $A$, $G_{IJ}$ and $\Bh^J$ general tensors built from the couplings $g^I$ and with $G_{IJ}=G_{(IJ)}$. As observed in \cite{Jack:2018oec,Pannell:2024sia,Pannell:2025ajf}, even the most general expansions of $A$ and $G_{IJ}$ cannot solve this equation at each order without factors of $1/\varepsilon$. The coefficients of $\Bh^J$ must therefore satisfy some of these equations on their own. These constraints are scheme independent, and the gradient conditions (\ref{eq:NAgrad}) and (\ref{eq:Agrad}) imply that the computed coefficients satisfy all of them.

In this appendix we summarise results for the multiscalar $\phi^4$ theory in $d=4-\varepsilon$ to seven loops. The constraints follow from enumerating the graphs allowed by the form of the couplings. The long-range $\phi^4$ theory in $2<d<4$ has the same couplings, so it is subject to the same constraints, although the coefficients that enter them differ.

There are 1747 constraints in $d=4$, and 311 more from extending the solution to $d=4-\varepsilon$, as summarised in Table \ref{tab:phi4constraints}. The constraints, and the coefficients of the short-range beta function through seven loops from \cite{Schnetz:2022nsc} are available in \cite{Stergiou:phi4GraphAtlas}.

\begin{table}[ht]
    \centering
    \begin{tabular}{|c|c|c|c|c|c|}
    \hline
        Loop order & 1--3 & 4 & 5 & 6 & 7 \\ \hline
        Constraints on $\Bh^I$ & 0 & 4 & $37+4^{\star}$ & $225+41^\star$ & $1481+266^\star$ \\ \hline
    \end{tabular}
    \caption{Number of constraints imposed by gradient flow on $\Bh^I$ in $\phi^4$ theory, by loop order through seven loops. An entry $n+m^\star$ gives the number $n$ of linearly independent constraints in $d=4$ and the number $m$ of additional constraints in $d=4-\varepsilon$.}
    \label{tab:phi4constraints}
\end{table}

For the short-range theory, the constraints show that $\upsilon$ is needed for the gradient property. Reference \cite{Pannell:2024sia} found that $\bh^I$ violates the constraints from six loops, while $\Bh^I$ might satisfy them. That reference did not compute $\upsilon$. We computed all 163 contributions to $\upsilon$ through seven loops, following \cite{Fortin:2012hn}, and the coefficients are also available in \cite{Stergiou:phi4GraphAtlas}. All 2058 constraints are satisfied by $\Bh^I$, in agreement with our general proof in this paper.

\section{Weyl consistency conditions in \texorpdfstring{$d$}{d} dimensions}\label{app:consistency}
In this appendix we derive the conditions (\ref{eq:d4-WZ-a})--(\ref{eq:d4-WZ-curl}) in $d=4-\ve$, together with the scheme without $\nabla^2\sigma$ terms. We set $a_\mu=0$ and omit the variation of $a_\mu$ in (\ref{eq:localRG}), so that $D_\mu=\partial_\mu$ and no terms with $\tilde\rho_I$ appear. These flavour terms in (\ref{eq:d4-Shat}) and (\ref{eq:d4-WZ-a}) arise as at $d=4$ \cite{Jack:2013sha}. We write $\delta_\sigma$ for the variation of the metric and $\Delta_\sigma$ for the full operator in (\ref{eq:localRG}), and we multiply (\ref{eq:d4-anomaly}) by $16\pi^2\mu^\ve$.

\subsection{Weyl variations}
Under $\delta_\sigma\gamma_{\mu\nu}=2\sigma\gamma_{\mu\nu}$, the volume element and the curvatures vary as
\begin{equation}\label{eq:wz-weyl}
    \begin{gathered}
        \delta_\sigma\sqrt\gamma=d\lsp\sigma\sqrt\gamma\,,\qquad\delta_\sigma R=-2\sigma R-2(d-1)\nabla^2\sigma\,,\\
        \delta_\sigma E^{\mu\nu}=-4\sigma E^{\mu\nu}-(d-2)(\nabla^\mu\partial^\nu-\gamma^{\mu\nu}\nabla^2)\sigma\,,\\
        \delta_\sigma\cE_4=-4\sigma\cE_4+8(d-3)E^{\mu\nu}\nabla_\mu\partial_\nu\sigma\,,
    \end{gathered}
\end{equation}
which at $d=4$ are (2.13) of \cite{Jack:2013sha}. The couplings vary by $\sigma\Bh^I$, so that
\begin{equation}\label{eq:wz-dg}
    \Delta_\sigma\partial_\mu g^I=\partial_\mu\sigma\lsp\Bh^I+\sigma(\partial_J\Bh^I)\partial_\mu g^J\,.
\end{equation}
The Laplacian of a fixed scalar $X$ varies as $\delta_\sigma\nabla^2X=-2\sigma\nabla^2X+(d-2)\partial_\mu\sigma\lsp\partial^\mu X$, and adding $\nabla^2(\sigma\Bh^I)$ gives
\begin{equation}\label{eq:wz-psi}
    \Delta_\sigma\nabla^2g^I=\Bh^I\nabla^2\sigma+2\lsp\Psi_J{}^I\partial_\mu\sigma\lsp\partial^\mu g^J+\text{terms proportional to }\sigma\,,
\end{equation}
where $\Psi_J{}^I=\frac{d-2}{2}\delta_J{}^I+\partial_J\Bh^I$. Without the flavour terms, (\ref{eq:d4-Shat}) reads $\widehat{\sS}_{IJ}=\sS_{IK}\Psi_J{}^K+\sT_{IJK}\Bh^K$.

\subsection{The scheme}
In the basis of (2.4) of \cite{Jack:2013sha}, a general scheme also has the terms
\begin{equation}\label{eq:wz-laplacian}
    -\int\dd v\,\nabla^2\sigma\Big(\tfrac16\sD R+\sU_I\nabla^2g^I+\sV_{IJ}\partial_\mu g^I\partial^\mu g^J\Big)\,.
\end{equation}
We remove them by adding to $16\pi^2\mu^\ve\ln Z$ the finite local functional
\begin{equation}\label{eq:wz-curvature-ct}
    \int\dd v\,R\lsp\Big(p\lsp R+p_I\nabla^2g^I+p_{IJ}\partial_\mu g^I\partial^\mu g^J\Big)\,,
\end{equation}
with regular coefficients. Varying the explicit factor of $R$ gives $-2(d-1)\nabla^2\sigma$ times the bracket, so $p_I$ and $p_{IJ}$ can cancel $\sU_I$ and $\sV_{IJ}$. Varying both factors of $R$ in $p\lsp R^2$ gives $-4(d-1)\lsp p\lsp R\nabla^2\sigma$, varying $\nabla^2g^I$ gives $p_I\Bh^IR\nabla^2\sigma$, and $p$ cancels these two terms together with $\sD$. In the resulting scheme the conditions take the form (\ref{eq:d4-WZ-a})--(\ref{eq:d4-WZ-curl}), and no $\nabla^2\sigma$ term enters the sphere evaluation of section \ref{sec:fourd}. The potential and metric (\ref{eq:AgradDef}) are unaffected, since $r\partial_r+\ve$ annihilates local terms on the sphere.

Every term in the Weyl variation of (\ref{eq:wz-curvature-ct}) contains $R$ or $\nabla^2\sigma$, so none multiplies the structures of $\sA$, $\sA_{IJ}$, $\sG_{IJ}$, $\sw_I$, $\sS_{IJ}$ and $\sT_{IJK}$, which are therefore unchanged. The functional shifts $\sB$, $\sE_I$, $\sH_I$ and $\sF_{IJ}$, and the conditions refer to the shifted coefficients. In a general scheme, (\ref{eq:d4-WZ-EH}) also contains $\sU_I$ and $\sV_{IJ}$, as in (2.16) of \cite{Jack:2013sha} at $d=4$.

\subsection{Derivation of the conditions}
Commutativity of Weyl rescalings requires $\Delta_{\sigma'}\mathcal{A}_\sigma-\Delta_\sigma\mathcal{A}_{\sigma'}=0$. Terms without derivatives of either Weyl factor cancel, and after integration by parts the others take the form
\begin{equation}\label{eq:wz-commutator}
    \int\dd v\,(\sigma\partial_\mu\sigma'-\sigma'\partial_\mu\sigma)Z^\mu+\int\dd v\,(\partial_\mu\sigma\nabla^2\sigma'-\partial_\mu\sigma'\nabla^2\sigma)V^\mu=0\,,
\end{equation}
where the vectors $Z^\mu$ and $V^\mu$ are local in the sources. As operators on arbitrary $\sigma'$, the two integrals have orders one and three, so $V^\mu=0$ and then $Z^\mu=0$. The reduction uses three identities, for a scalar $X$, a symmetric tensor $X^{\mu\nu}$ and an antisymmetric tensor $Y^{\mu\nu}$,
\begin{equation}\label{eq:wz-reductions}
    \begin{aligned}
        \int\dd v\,X(\sigma\nabla^2\sigma'-\sigma'\nabla^2\sigma)&=-\int\dd v\,(\sigma\partial_\mu\sigma'-\sigma'\partial_\mu\sigma)\partial^\mu X\,,\\
        \int\dd v\,X^{\mu\nu}(\sigma\nabla_\mu\partial_\nu\sigma'-\sigma'\nabla_\mu\partial_\nu\sigma)&=-\int\dd v\,(\sigma\partial_\nu\sigma'-\sigma'\partial_\nu\sigma)\nabla_\mu X^{\mu\nu}\,,\\
        \int\dd v\,Y^{\mu\nu}\partial_\mu\sigma\lsp\partial_\nu\sigma'&=-\tfrac12\int\dd v\,(\sigma\partial_\nu\sigma'-\sigma'\partial_\nu\sigma)\nabla_\mu Y^{\mu\nu}\,.
    \end{aligned}
\end{equation}
The first two follow by differentiating $\sigma\partial\sigma'-\sigma'\partial\sigma$, and the third by integrating by parts on either Weyl factor and averaging.

\paragraph{The conditions for $\sH_I$ and $\sE_I$.} Terms with $\partial_\mu\sigma$ and two derivatives of $\sigma'$ come from varying $R$, $\nabla^2g^I$ and $E^{\mu\nu}$ in the terms with $\sH_I$, $\sS_{IJ}$ and $\sw_I$. By (\ref{eq:wz-weyl}) and (\ref{eq:wz-psi}), they are
\begin{equation}\label{eq:wz-three-derivative}
    \int\dd v\,\partial_\mu\sigma\Big[2\Big(\tfrac{d-1}{3}\lsp\sH_I-\sS_{IJ}\Bh^J\Big)\partial^\mu g^I\nabla^2\sigma'+2(d-2)\sw_I\partial_\nu g^I(\nabla^\mu\partial^\nu-\gamma^{\mu\nu}\nabla^2)\sigma'\Big]\,.
\end{equation}
After antisymmetrisation, the term with $\sw_I$ is a total derivative of first derivatives, because
\begin{equation}\label{eq:wz-w-total}
    \partial_\mu\sigma(\nabla^\mu\partial^\nu-\gamma^{\mu\nu}\nabla^2)\sigma'-(\sigma\leftrightarrow\sigma')=\nabla_\mu(\partial^\mu\sigma\lsp\partial^\nu\sigma'-\partial^\mu\sigma'\lsp\partial^\nu\sigma)\,.
\end{equation}
It therefore contributes only to $Z^\mu$, which leaves $V^\mu=2(\frac{d-1}{3}\sH_I-\sS_{IJ}\Bh^J)\partial^\mu g^I$. Its vanishing is the second relation in (\ref{eq:d4-WZ-EH}).

The first relation comes from varying $R$ in the term with $\sE_I$ and $\nabla^2g^I$ in the term with $\sA_{IJ}$, which gives
\begin{equation}
    \int\dd v\,\sigma\nabla^2\sigma'\Big(\tfrac{d-1}{3}\lsp\sE_I+\sA_{IJ}\Bh^J\Big)\nabla^2g^I\,.
\end{equation}
By the first reduction in (\ref{eq:wz-reductions}), this contributes $-(\frac{d-1}{3}\sE_I+\sA_{IJ}\Bh^J)\partial^\mu\nabla^2g^I$ to $Z^\mu$. No other term in the anomaly produces a third derivative of the couplings, so this coefficient vanishes, which is the first relation in (\ref{eq:d4-WZ-EH}).

\paragraph{The curl relation.} Terms with one derivative on each Weyl factor arise from varying $\nabla^2g^I$ through $\Psi$ in the term with $\sS_{IJ}$, from varying $\partial^\nu g^J\partial_\nu g^K$ through (\ref{eq:wz-dg}) in the term with $\sT_{IJK}$, and from the term with $\sw_I$ after integrating (\ref{eq:wz-w-total}) by parts. Terms proportional to $\partial_\mu\sigma\lsp\partial^\mu\sigma'$ cancel, and the remainder is
\begin{equation}\label{eq:wz-curl-terms}
    -2\int\dd v\,\partial_\mu\sigma\lsp\partial_\nu\sigma'\lsp\partial^\mu g^I\partial^\nu g^J\Big[(d-2)(\partial_I\sw_J-\partial_J\sw_I)+2(\widehat\sS_{IJ}-\widehat\sS_{JI})\Big]\,.
\end{equation}
The coefficient of $\partial_\mu\sigma\lsp\partial_\nu\sigma'$ is antisymmetric in $\mu$ and $\nu$, so the third reduction in (\ref{eq:wz-reductions}) puts its divergence in $Z^\nu$, as $\partial_\mu g^I\nabla^\mu\partial^\nu g^J$ with an antisymmetric coupling coefficient. The other contributions to this structure, from $\sG_{IJ}$, $\sF_{IJ}$ and the omitted terms through the first two reductions, have symmetric coefficients. The bracket in (\ref{eq:wz-curl-terms}) therefore vanishes, which is (\ref{eq:d4-WZ-curl}).

\paragraph{The Einstein relation.} The terms in $Z^\mu$ proportional to $E^{\mu\nu}\partial_\nu g^I$ take the form
\begin{equation}\label{eq:wz-einstein}
    2\int\dd v\,(\sigma\partial_\mu\sigma'-\sigma'\partial_\mu\sigma)X_IE^{\mu\nu}\partial_\nu g^I\,,
\end{equation}
with three contributions to $X_I$. First, varying $\cE_4$ gives $-2(d-3)\int\dd v\,\sigma\sA E^{\mu\nu}\nabla_\mu\partial_\nu\sigma'$, which contributes $(d-3)\partial_I\sA$ after integrating by parts with $\nabla_\mu E^{\mu\nu}=0$ and antisymmetrising. Second, varying either $\partial g$ in the term with $\sG_{IJ}$ contributes $-\sG_{IJ}\Bh^J$. Third, the density of the term with $\sw_I$ has Weyl weight $d-4=-\ve$, since the measure has weight $d$, $E^{\mu\nu}$ has weight $-4$ and lower-index derivatives have weight zero. The variation of the couplings acts on $\sw_I\partial_\nu g^I$ through $(\stL_{\Bh}\sw)_I=\Bh^J\partial_J\sw_I+(\partial_I\Bh^J)\sw_J$, up to a term $\partial_\nu\sigma'\sw_J\Bh^J$ that cancels by symmetry after contraction with $\partial_\mu\sigma E^{\mu\nu}$. Hence $X_I=(d-3)\partial_I\sA-\sG_{IJ}\Bh^J+(\stL_{\Bh}\sw)_I-\ve\lsp\sw_I$, and $X_I=0$ is (\ref{eq:d4-WZ-a}) without its flavour terms.

\printbibliography

\end{document}